\documentclass[pre,twocolumn,showpacs,showkeys]{revtex4-2}
\usepackage{color}
\usepackage{graphicx,hyperref,amsmath,soul}
\usepackage{mathtools,bm}
\usepackage{graphicx}
\usepackage{subfigure}
\usepackage{float}

\begin{document}
\title{Disparity Driven Heterogeneous Nucleation in Finite-Size Adaptive Networks}
\author{Akash Yadav$^1$, Jan Fialkowski$^{2,3}$, Rico Berner$^{4}$,  V. K. Chandrasekar$^5$ and  D. V. Senthilkumar$^{1}$}
\affiliation{ $^1$School of Physics, Indian Institute of Science Education and Research, Thiruvananthapuram-695551, Kerala, India\\
	$^2$ Complexity Science Hub Vienna, Josefstädter Straße 39, 1080 Vienna, Austria\\
	$^3$ Center for Medical Data Science, Medical University Vienna, Spitalgasse 23, 1090 Vienna, Austria\\
	$^4$Department of Physics, Humboldt-Universit\"at zu Berlin, Newtonstraße 15, 12489 Berlin, Germany\\
	$^5$Centre for Nonlinear Science \& Engineering, School of
	Electrical \& Electronics Engineering, SASTRA Deemed University,
	Thanjavur-613401, Tamil Nadu, India
}

\date{\today}
\begin{abstract}
	Phase transitions are crucial  in shaping the collective dynamics of a broad spectrum of natural systems across disciplines.  Here,
	we report two distinct heterogeneous nucleation facilitating  single-step and  multi-step phase  transitions to global synchronization in
	a finite-size  adaptive network due to the  trade-off between  time scale adaptation and coupling strength disparities.  Specifically,
	small intracluster nucleations coalesce  either at the population interface or within the populations resulting in the two distinct phase transitions
	depending on the degree of the disparities.  We find that the coupling strength disparity largely controls the nature of phase transition in the phase diagram
	irrespective of the adaptation disparity.  We provide a mesoscopic description for the cluster dynamics using the  collective coordinates approach that
	brilliantly captures the multicluster dynamics among the populations leading to distinct phase transitions.
	Further, we also deduce the upper bound for the coupling strength for the existence of two intraclusters explicitly in terms of adaptation
	and coupling strength disparities.
	These insights may have implications across domains ranging from neurological disorders to segregation dynamics in social networks.
\end{abstract}
\maketitle
\textit{Introduction.}
Complex systems ~\cite{sbvl2006,strogatz2001,newman2003}, characterized by their intricate interconnections, often exhibit transitions from incoherence to coherence ~\cite{boccaletti2018,dorfler2013,shahal2020}. Phase transitions are observed in several physical phenomena including crystallization and ferromagnetic transition.
Beyond the physical systems, phase transitions in complex networks can shed  more light on intriguing biological, ecological, and social problems such as population collapse
and species extinction~\cite{bdmp2011}, polarization in society~\cite{conover2011}, and crashes in financial markets~\cite{levy2008stock}. Particularly,
transitions from incoherence to synchronization have been extensively studied employing various network topologies of real-world
relevance~\cite{alex2008,gjya2007,dskp2023}. Traditionally, most studies have been primarily concentrated on static networks. However, a large class of real-world networks
co-evolve with their dynamical states and adapt to the prevailing environments.  For instance, from  intricate dynamics of human brain~\cite{gerstner1996,cndy2008,rohr2019},
technological systems~\cite{waldrop2013,MORALES2021,brys2021},  and biological networks~\cite{PROULX2005,gtdc2006,irmg2011} to social
dynamics~\cite{hlkc2020,bfls2020,addc2015} adaptability manifests   in the entire  spectrum of complex networks across disciplines.

Recent interest in adaptive dynamical networks has surged due to their potential significance in addressing complex systems ~\cite{sjbr2023,jube2023}. Adaptively coupled phase
oscillator models are  paradigms for describing the interplay of function and structure in complex systems~\cite{berner2021}. Dynamical features such as frequency clusters~\cite{berner2019multi,thch2022},
solitary states~\cite{berner2020solitary}, recurrent synchronization~\cite{thiele2023asymmetric}, and heterogeneous nucleation~\cite{fial2023} are specific to adaptive dynamical networks~\cite{BERNER2023}. In addition to adaptivity,
large-scale systems consist of multiple populations exhibiting diverse and multi-scale behavior~\cite{girvan2002,newman2006,mucha2010}. Examples include dynamics of distinct
brain regions spanning multiple time scales~\cite{BETZEL201773},  ecological communities  exhibiting different time scales in response to  infectious diseases~\cite{grja2020,dbdh2013,qgfg2020},  and time scales of social ties formation also vary across communities~\cite{fljc2012,saramaki2015}.  These investigations also reveal that
time scale disparity plays a nontrivial role  in shaping their collective dynamics.
Studies have shown  that depending on various factors, a system may opt for different routes during phase transition. For instance, multiple nucleation pathways can unfold in
crystal formation, each involving distinct intermediate states ~\cite{guo2016,xu2021role}. Similarly, in the opinion formation dynamics on social networks, individuals can form a
cohesive community with consensus or can form fragmented structures known as echo chambers~\cite{evans2018,toth2021}. The ability to manipulate the transition pathways holds immense importance as it allows to steer the system through the appropriate intermediate states under favorable conditions.

In this work, we consider a finite-size adaptive network comprised of two populations with time scale adaptation disparity and coupling strength disparity without
any quenched disorder.  We observe two distinct nucleation. In the first scenario, we find a single large interpopulation frequency cluster emerges at the population interface due to the coalescence of small intrapopulation clusters.
The single large inter-frequency-cluster  eventually enlarges to the system size as a function of the coupling strength facilitating a multi-step transition to
global synchronization.  In sharp contrast, intrapopulation clusters nucleate and coalesce together to manifest two completely entrained  intrapopulation clusters
as a function of the coupling strength for a strong intrapopulation adaptation rate. Finally, the two intraclusters merge together for a large coupling strength
resulting in a single-step transition to synchronization.  Recently, similar heterogeneous nucleation resulting in multi- and single-step transitions are reported
to be facilitated by distinct quenched disorders~\cite{fial2023}.   A strong interpopulation coupling strength always favors nucleation of inter-frequency-clusters
leading to multi-step transition even with a strong intrapopulation  adaptation rate.  Similarly, a strong intrapopulation coupling strength always favors nucleation
of intra-frequency-clusters facilitating single-step transition even with a  strong interpopulation  adaptation rate. These results reveal that  the disparity in the
coupling strength determines the nature of nucleation leading to distinct synchronization transition.  We  analytically  deduce the  macroscopic evolution equations
for the cluster dynamics using the collective coordinates framework~\cite{gwga2015}  and show that the latter corroborates the simulation results.  Further, we also
deduce the upper bound for the coupling strength for the existence of two intraclusters explicitly, at which the abrupt single-step transition manifests,
in terms of adaptation and coupling strength disparity parameters.\\
\indent\textit{The Model.}
We consider a system of $N$  globally coupled phase oscillators  with adaptive coupling represented as
\begin{subequations}\label{eq:model}
	\begin{eqnarray}
		\frac{d\phi_{i}^{\eta}}{dt} = \omega_{i}^{\eta} - \frac{1}{N} \sum _{\eta'} \sigma_{\eta \eta'} \sum_{j=1}^{N_{\eta'}} \kappa_{ij}^{\eta\eta'} \sin{(\phi_{i}^{\eta}-\phi_{j}^{\eta'})},\label{eq1a}\\
		\frac{d\kappa_{ij}^{\eta\eta'}}{dt} = -\varepsilon_{\eta\eta'} \left [  \kappa_{ij}^{\eta\eta'} + \sin{(\phi_{i}^{\eta}-\phi_{j}^{\eta'}+\beta)}   \right ],\label{eq1b}
	\end{eqnarray}
	\label{eq:phase}
\end{subequations}
where $\omega_i^{\eta}$  and $\phi_i^\eta(t)$  are the natural frequency and  the phase of the  $i^{th}$ oscillator ($i= 1,2, \ldots ,N_\eta$) in the $\eta^{th}$ population, respectively.
Here, we consider two populations $\eta=\{A,B\}$. The coupling weights $\kappa_{ij}^{\eta\eta'}(t) \in [-1,1]$ co-evolve with the phases of  the oscillators,
$\sigma_{\eta\eta'}$ is the coupling strength,  $\varepsilon_{\eta\eta'}$ is  the time-scale parameter determining the adaptation rate of the  coupling weights, and the parameter $\beta$
accounts for different adaptation rules~\cite{atat2009,atat2011}.  We have fixed $\beta= -0.53\pi$, close to the symmetric  rule~\cite{fial2023}, which is also referred to as Hebbian adaptation rule.

In this model, Eq.~(\ref{eq1a}) governs the phase dynamics, which is a generalized version of the Kuramoto model with adaptive coupling.  The coupling matrix $\kappa(t)$ characterizes the network topology, and the dynamics of coupling weights is governed by Eq.~(\ref{eq1b}). To investigate the nontrivial effect of disparities among populations, we consider
equally sized populations with homogeneous(heterogeneous) coupling strength and time scale  adaptation among intra-(inter-) population.
Further, the interactions are chosen to be symmetric with
$(\sigma,\varepsilon)_{AB} = (\sigma,\varepsilon)_{BA}$ and  $(\sigma,\varepsilon)_{AA} = (\sigma,\varepsilon)_{BB}$.
The disparity between intra- and interpopulations coupling
strengths is governed by $\sigma_{AA} = \sigma(1+\Lambda_{\sigma})$ and $\sigma_{AB} = \sigma(1-\Lambda_{\sigma})$, where $\Lambda_{\sigma}$ is the
coupling strength disparity parameter, and $\sigma$ is the control parameter.  Analogously, the disparity in time scales is governed by $\varepsilon_{AA} = \varepsilon(1+\Lambda_{\varepsilon})$ and
$\varepsilon_{AB} = \varepsilon(1-\Lambda_{\varepsilon})$,  where $\Lambda_{\varepsilon}$ is the adapation disparity parameter and $\varepsilon$ is set to $0.01$.
When $\Lambda_\sigma=\Lambda_\varepsilon=0$, there is neither coupling  disparity nor adaptation disparity between $A$ and $B$. Intrapopulation coupling strengths
and adapations are  larger for $(\Lambda_\sigma, \Lambda_\varepsilon)>0$  and smaller for $(\Lambda_\sigma, \Lambda_\varepsilon)<0$.
Note that rescaling of time $t$ with the transformations $\omega_i\rightarrow \omega'_i/\tau$, $\sigma\rightarrow \sigma'/\tau$, and $\varepsilon\rightarrow \varepsilon'/\tau$ recovers the dynamics of the model, where $\tau$ is a constant factor.

\begin{figure}[ht]
	\includegraphics[width=0.5\textwidth]{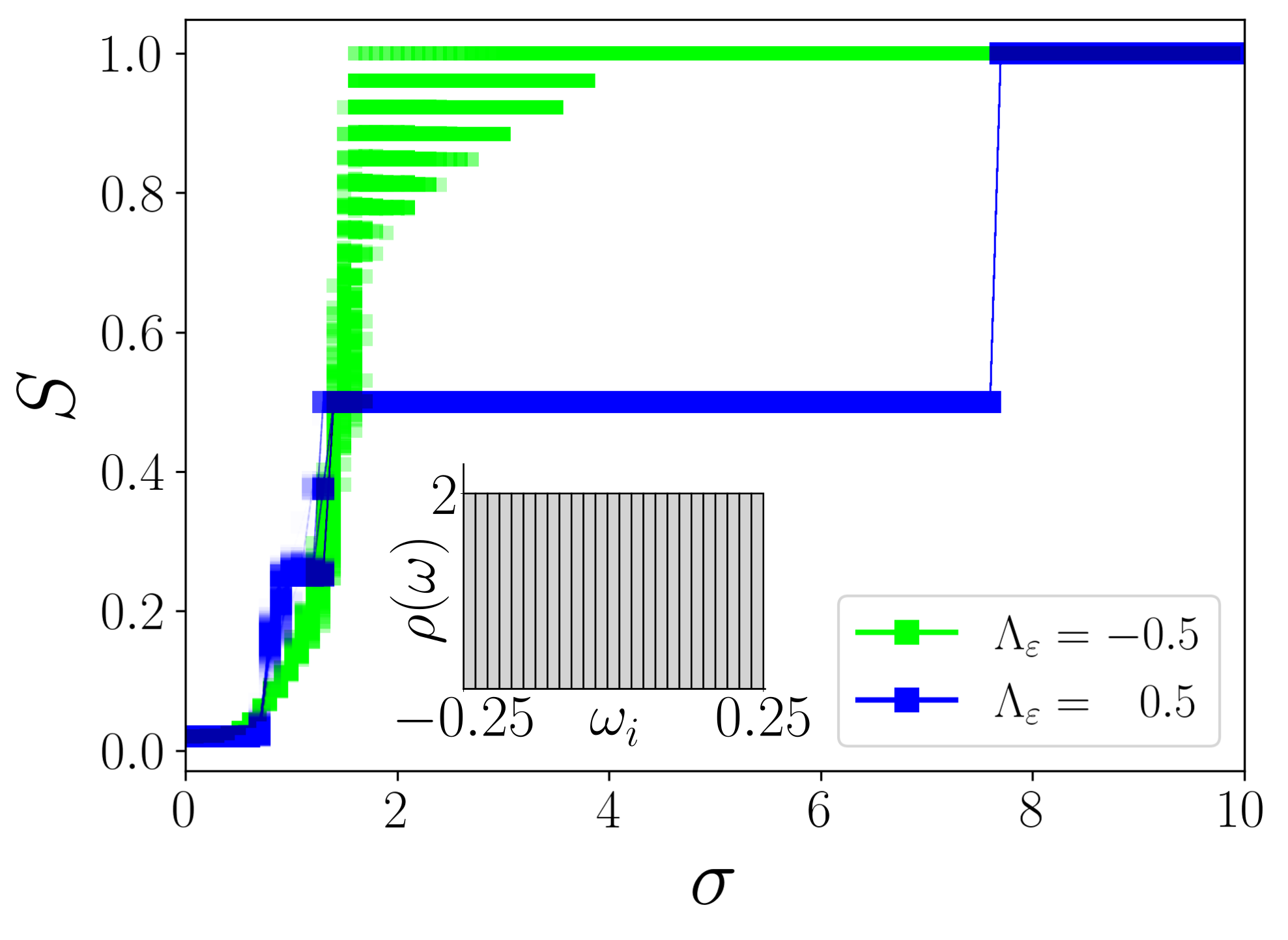}
	\caption{Synchronization transition  of  the system of  $N=50$ globally coupled  phase oscillators (\ref{eq:phase}) for  500 realizations.  The coupling  strength disparity  is fixed as
		$\Lambda_\sigma=0$. The system undergoes a multi-step transition for $\Lambda_{\varepsilon}=-0.5$, whereas for $\Lambda_{\varepsilon}=0.5$ system follows
		a single-step transition to synchrony.  Inset  illustrates that  the natural frequencies of the oscillators are drawn from a uniform distribution in the
		range of $[-0.25, 0.25]$.  We have fixed $\delta=0.001$.  Other parameters are $\beta=-0.53\pi$ and  $\varepsilon=0.01$. See the main text for more details.}
	\label{fig:transition}
\end{figure}
\textit{Results.}
The system of  $N = 50$  adaptively coupled phase oscillators   (\ref{eq:phase}) are numerically solved using the Runge-Kutta 4th order integration scheme.
We assign oscillators in the range $i=1,\hdots, N/2$ to the first population, and oscillators in the range $N/2 +1,\hdots, N$ are assigned to the second population.
The oscillators  in each population are sorted in the increasing order of  their natural frequencies. $\phi_i^\eta$'s are chosen randomly from the interval
$ [0,2\pi)$.
We have fixed $\kappa_{ij}^{\eta\eta'}(0)=0~\forall~i, j$, $\varepsilon=0.01$ and $\beta = -0.53\pi$.

\begin{figure}[ht]
	\includegraphics[width=0.5\textwidth]{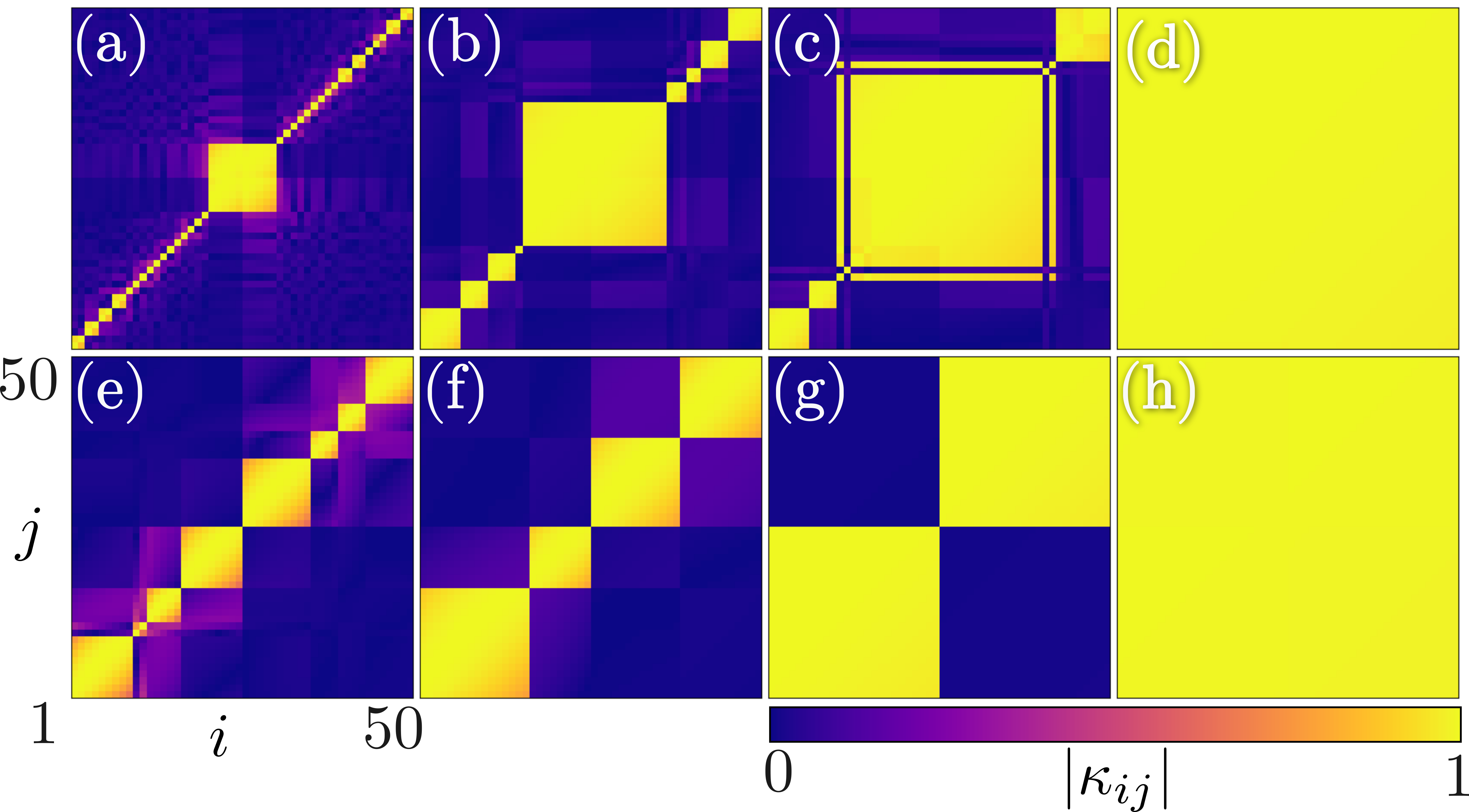}
	\caption[short]{The evolution of coupling weights with coupling strength $\sigma$ in the absence of the coupling strength disparity $\Lambda_\sigma=0$.
		(a)-(d) $\Lambda_\varepsilon= -0.5$, and (e)-(h) $\Lambda_\varepsilon=0.5$.
		The values of $\sigma$ are (a) $0.7$, (b) $1.2$, (c) $1.4$, (d) $3.9$, (e) $0.75$, (f) $0.95$, (g) $5.0$,  and (h) $7.7$. Other parameters are $\beta=-0.53\pi$ and  $\varepsilon=0.01$. See the main text for more details.}
	\label{fig:coupling_weights}
\end{figure}
Initially, the system is in an incoherent state for  $\sigma=0$, and small coupling strengths. As the coupling strength increases, the oscillators self-organize to global synchronization.
The oscillators with larger $\kappa_{ij}^{\eta\eta'}$  tend to synchronize  more quickly than that with smaller $\kappa_{ij}^{\eta\eta'}$.
In the following, we show that the system undergoes qualitatively  two distinct routes to synchrony  as governed by the disparities among the populations.

We employ the synchronization index $S$~\cite{fial2023}, that quantify the  degree of coherence in the network, represented as
\begin{equation}\label{eq:sync_index}
	S = \frac{1}{N^2} \sum_{\eta,\eta'} \sum_{i=1}^{N_\eta}\sum_{j=1}^{N_{\eta'}} s_{ij}^{\eta\eta'},
\end{equation}
where $s_{ij}^{\eta\eta'}$ measures the pairwise frequency synchronization between $i^{th}$ and $j^{th}$ oscillator defined as
\begin{equation}\label{eq:sync_def}
	s_{ij}^{\eta\eta'} =
	\begin{cases}
		1, & \text{if }  \lvert \langle \dot{\phi}_i^\eta \rangle -\langle \dot{\phi}_j^{\eta'} \rangle\rvert \leq \delta, \\
		0, & \text{if } \lvert \langle \dot{\phi}_i^\eta \rangle -\langle \dot{\phi}_j^{\eta'} \rangle\rvert > \delta,
	\end{cases}
\end{equation}
where  $\delta$ is a predefined threshold and $\langle \dot{\phi}_i^\eta \rangle = \lim_{T \to \infty} (1/T)\int_{T_0}^{T_0+T} \dot{\phi}_i^\eta dt $
is the mean phase velocity of the  $i^{th}$ oscillator  calculated after a  large transient  $T_0$. For $S=1$, the system is completely synchronized,
whereas $S=1/N$ indicates complete incoherence.

First, we discuss the phase transition facilitated  solely by adaptation disparity $\Lambda_\varepsilon$ for $\Lambda_\sigma=0$.
The network (\ref{eq:phase}) exhibits a multi-step transition to global synchronization  as a function of $\sigma$ for $\Lambda_\varepsilon=-0.5$
as depicted in  Fig.~\ref{fig:transition}  (green/light grey squares).  The dynamics of the coupling weights  $\kappa_{ij}^{\eta\eta'}(t)$ unveil  crucial insights on
the underlying mechanism for such a transition as they co-evolve with the  phases of the oscillators.  The snapshots of  $\kappa_{ij}^{\eta\eta'}(t)$
are plotted in Figs.~\ref{fig:coupling_weights}(a)-(d)  for $\Lambda_\varepsilon=-0.5$.  It is evident that  a single frequency cluster, entrained by oscillators from both populations,
emerges at the interface of the populations, which grows with an increase in the coupling strength due to  the  subsequent entrainment of smaller intrapopulation clusters.
This process of continuous integration of smaller clusters (compare Figs.~\ref{fig:coupling_weights}(a)-(c)), as a function of  $\sigma$, leads to a staggered
rise in the degree of synchronization $S$ (see  Fig.~\ref{fig:transition})  and   eventually  results in global synchronization (see Fig.~\ref{fig:coupling_weights}(d)).
Note that, for $\Lambda_\varepsilon<0$,  the interpopulation  adaptation rate is larger than intrapopulation adaptation, that is $\varepsilon_{AB}>\varepsilon_{AA}$, and hence
oscillators from different populations  rapidly entrained to form nucleations, which grow as a single cluster as observed in Figs.~\ref{fig:coupling_weights}(a)-(d).

\begin{figure*}[ht]
	\includegraphics[width=\textwidth]{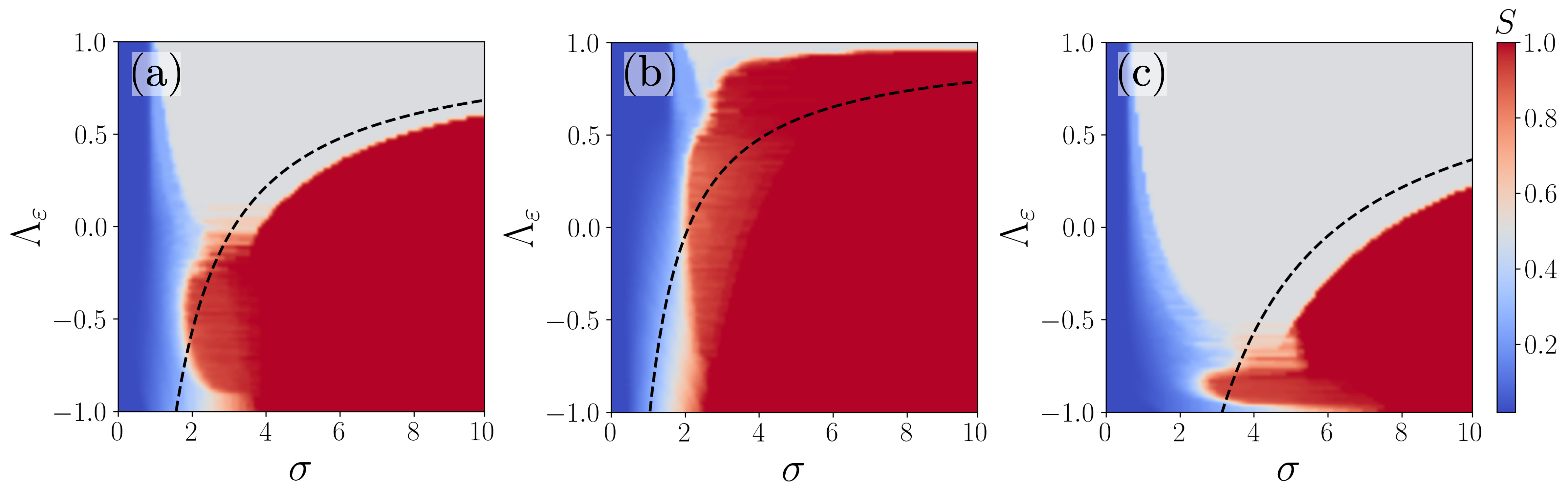}
	\caption{Heat maps of synchronization index $S$, for $10$ different realizations, in the ($\sigma, \Lambda_{\varepsilon}$) parameter space. (a) $\Lambda_{\sigma} =0$,
		(b) $\Lambda_{\sigma} = -0.5$, and  (c)  $\Lambda_{\sigma} =0.5$. The dashed black curve is  the analytical estimate for the upper bound of the two cluster state. Other parameters are $\beta=-0.53\pi$ and  $\varepsilon=0.01$. See the main text for more details.}
	\label{fig:phase_plots}
\end{figure*}

In stark contrast, for  $\Lambda_\varepsilon>0$,  the network exhibits a single-step transition to global synchronization as corroborated by
$S$  in Fig.~\ref{fig:transition} for  $\Lambda_\varepsilon=0.5$ (blue/dark grey squares).  Again, $\kappa_{ij}^{\eta\eta'}(t)$
uncover the underlying mechanism for such an abrupt  synchronization transition. The snapshots of $\kappa_{ij}^{\eta\eta'}(t)$   for
$\Lambda_\varepsilon=0.5$ are depicted in Figs.~\ref{fig:coupling_weights}(e)-(h).  Coexisting intrapopulation clusters emerge and subsequently enlarge by the entrainment of
smaller intrapopulation clusters   (see Figs.~\ref{fig:coupling_weights}(e) and (f)) manifesting entrained subpopulations for a large $\sigma$ (see Fig.~\ref{fig:coupling_weights}(g)).
Finally, the completely entrained intrapopulation clusters coalesce together resulting in  global synchronization   (Fig.~\ref{fig:coupling_weights}(h))
as corroborated by an abrupt jump in the synchronization index  to $S=1$ (see Fig.~\ref{fig:transition} for  $\Lambda_\varepsilon = 0.5$).
Note that the intrapopulation  adaptation is larger than interpopulation adaptation  for $\Lambda_\varepsilon>0$, that is  $\varepsilon_{AA}>\varepsilon_{AB}$,  and consequently,
nucleations are formed within the populations resulting in  coexisting intrapopulation clusters.  Thus, one can reinforce a particular route to  the phase transition
by tuning the time scale of  adaptation.\\
\indent Heat maps of $S$ are depicted in the $(\sigma, \Lambda_\varepsilon)$ parameter space in Figs.~\ref{fig:phase_plots}(a)-(c) for three distinct $\Lambda_\sigma$.
For  $\Lambda_\sigma=0$,  clearly, there is a single-step(multi-step)  transition for $\Lambda_\varepsilon>0(<0)$ (see Fig.~\ref{fig:phase_plots}(a)) as  discussed above.
Nevertheless, the effect of trade-off between  $\Lambda_\sigma$ and $\Lambda_\varepsilon$ is evident from Figs.~\ref{fig:phase_plots}(b) and (c) depicted for
$\Lambda_\sigma=-0.5$ and $0.5$, respectively.
interpopulation coupling strength ($\sigma_{AB}$) is larger than that of intrapopulation ($\sigma_{AA}$)  in Fig.~\ref{fig:phase_plots}(b)
for $\Lambda_\sigma=-0.5$.  interpopulation adaption ($\varepsilon_{AB}$) is  also larger in the range of $\Lambda_\varepsilon\in[-1, 0.0)$ resulting in a multi-step transition.
Despite  the intrapopulation adaption ($\varepsilon_{AA}$) is larger than $\varepsilon_{AB}$, a strong $\sigma_{AA}$
facilitates multi-step transition in the range of $\Lambda_\varepsilon\in(0, 0.7)$ in Fig.~\ref{fig:phase_plots}(b).  This elucidates that the coupling disparity ($\Lambda_\sigma$)
dominates the adaptation disparity ($\Lambda_\varepsilon$)  in facilitating phase transitions.  However,  when  $\Lambda_\varepsilon\rightarrow 1$ $\varepsilon_{AB}\rightarrow 0$,
and hence a  large $\varepsilon_{AA}$ facilitates nucleation of  clusters within populations  resulting  in entrained intrapopulations and  eventually a single-step  transition
despite a large $\sigma_{AB}$ in the range of  $\Lambda_\varepsilon\in(0.7,0.9)$.  Further,  $\varepsilon_{AB}\approx 0$ when $\Lambda_\varepsilon\approx 1$ and consequently
only two-cluster state manifest without any global synchronization.
Both  $(\varepsilon, \sigma)_{AA}>(\varepsilon, \sigma)_{AB}$ facilitate distinct nucleation
within intrapopulations and  a single-step transition  in the range of $\Lambda_\varepsilon\in(0, 1)$  in  Fig.~\ref{fig:phase_plots}(c) for  $\Lambda_\sigma=0.5$.
Now,  again a large $\sigma_{AA}$ leads to a single-step transition even in the range of $\Lambda_\varepsilon\in(0, -0.5)$ (see Fig.~\ref{fig:phase_plots}(c)),
where interpopulation adaptation is larger, which reinforces that the coupling disparity ($\Lambda_\sigma$)  dominates the adaptation disparity.
However, $\varepsilon_{AB}>>\varepsilon_{AA}$  for $\Lambda_\varepsilon\in(-0.5, -1)$ and hence
a single large nucleation manifests at the population interface leading to a multi-step transition.\\

\indent\textit{Mesoscopic Dynamics.}
Being evident that the transition from asynchrony  to global synchronization involves the nucleation and merging of frequency clusters,
the dynamics of the full system can be captured by cluster level description. We employ a
collective coordinate approach to analyze the synchronization of intrapopulation clusters of distinct populations~\cite{gwga2015,hancock2018,smith2020,fial2023}.
The dynamical variables corresponding to the phase and coupling weights of an oscillator can be expressed in terms of collective coordinates
$\phi_{i,\mu}^{\eta}(t) $ and $\kappa_{\mu\nu}^{\eta\eta'}(t)$ with the ansatz,
\begin{subequations}
	\begin{eqnarray}
		\phi_{i}^{\eta}  \approx  \phi_{i,\mu}^{\eta} & = &  \Theta_{\mu}^{\eta}(t)(\omega_i^\eta-\Omega_\mu^\eta) + f_{\mu}^\eta(t), \\
		\kappa_{ij}^{\eta\eta'}(t)  & \approx & \kappa_{\mu\nu}^{\eta\eta'} (t).
	\end{eqnarray}
\end{subequations}

We consider there are $N_c^\eta$ number of clusters in the $\eta$-th population, and clusters are described by indices $\mu,\nu$.
The term $\Theta_\mu^\eta(\omega_i^\eta - \Omega_\mu^\eta)$ describes the  frequency drift  of $i^{th}$ oscillator within $\mu^{th}$ cluster in the population $\eta$, $\Omega_\mu^\eta$ is its
mean natural frequency and $f_\mu^\eta$ describes its collective phase. The intercluster ($\kappa_{\mu\nu}^{\eta\eta'}$) coupling weights  evolve in time.
The errors arising in describing the evolution of  the phase and coupling weights in terms of  collective coordinates  can be defined as
$E_{\phi_{i}^\eta} = \dot{\phi}_{i,\mu}^\eta - \dot{\phi}_{i}^\eta $ and $E_{\kappa_{ij}^{\eta\eta'}} = \dot{\kappa}_{\mu\nu}^{\eta\eta'}- \dot{\kappa}_{ij}^{\eta\eta'}$, respectively.
The evolution equations for the collective coordinates $(\Theta_\mu^\eta,f_\mu^\eta, \kappa_{\mu\nu}^{\eta\eta'})$ are obtained by requiring  the error
$\bm{E}= (E_{\phi_{1}^{\eta}}, \hdots, E_{\phi_{N_{\eta}}^{\eta}}, E_{\phi_{1}^{\eta'}}, \hdots, E_{\phi_{N_{\eta'}}^{\eta'}}, \hdots, E_{\kappa_{11}^{\eta\eta}}, \hdots, E_{\kappa_{N_\eta N_\eta}^{\eta\eta}}, \\E_{\kappa_{11}^{\eta'\eta'}},\hdots,  E_{\kappa_{N_{\eta'}N_{\eta'}}^{\eta'\eta'}}, \hdots)$ to be orthogonal to the manifold of the ansatz.
The frequency distribution  corresponding to $\mu^{th}$ cluster of  $\eta^{th}$ population can be expressed as $\rho_\mu^\eta(\omega) = 2/n_\mu^\eta \text{ for }  (\Omega_\mu^\eta -0.25 n_\mu^\eta \leq \omega \leq \Omega_\mu^\eta + 0.25 n_\mu^\eta), \text{and } 0 \text{ otherwise}$. Here $n_\mu^\eta$ is the ratio ($N_\mu^\eta/N$) of oscillators.
In the continuum limit,  the cluster order parameter $r_\mu^\eta = 1/N_\mu^\eta\left| \sum_{i\in C_\mu} e^{i\phi_{i,\mu}^\eta} \right| $ becomes $ r_\mu^\eta =(4/{n_\mu^\eta\Theta_\mu^\eta}) \sin{(n_\mu^\eta\Theta_\mu^\eta/4)}$ with  the variance of natural frequencies of the cluster as  $\xi_\mu^\eta = (n_\mu^\eta)^2/48$.
Consequently, the dynamics of the collective coordinates is governed by

%
\begin{subequations}\label{eq:collective}
	\begin{eqnarray}
		\dot{\Theta}_\mu^\eta  &=&   1  + \frac{1}{\xi_\mu^\eta \Theta_\mu^\eta} \left[\cos\left(\frac{n_\mu^\eta \Theta_\mu^\eta}{4}\right) -r_\mu^{\eta} \right] \nonumber \\ & \times & \left[ \sum_{\eta'} \sum_{\nu} n_\nu^{\eta'}  r_\nu^{\eta'} \sigma_{\eta\eta'}\kappa_{\mu\nu}^{\eta\eta'} \cos{(f_\mu^\eta - f_\nu^{\eta'})} \right],\\
		\dot{f}_\mu^\eta &=& \Omega_{\mu}^\eta - r_\mu^\eta \sum_{\eta'}  \sum_{\nu} n_\nu^{\eta'} r_\nu^{\eta'} \sigma_{\eta\eta'} \kappa_{\mu\nu}^{\eta\eta'} \sin{(f_\mu^\eta - f_\nu^{\eta'})},\\
		\dot{\kappa}_{\mu\nu}^{\eta\eta'} &=& -\varepsilon_{\eta\eta'} \left[ \kappa_{\mu\nu}^{\eta\eta'} + r_\mu^\eta r_\nu^{\eta'} \sin{(f_\mu^\eta - f_\nu^{\eta'} + \beta)} \right].
	\end{eqnarray}
\end{subequations}





\begin{figure}[ht]
	\centering
	\includegraphics[width=0.48\textwidth]{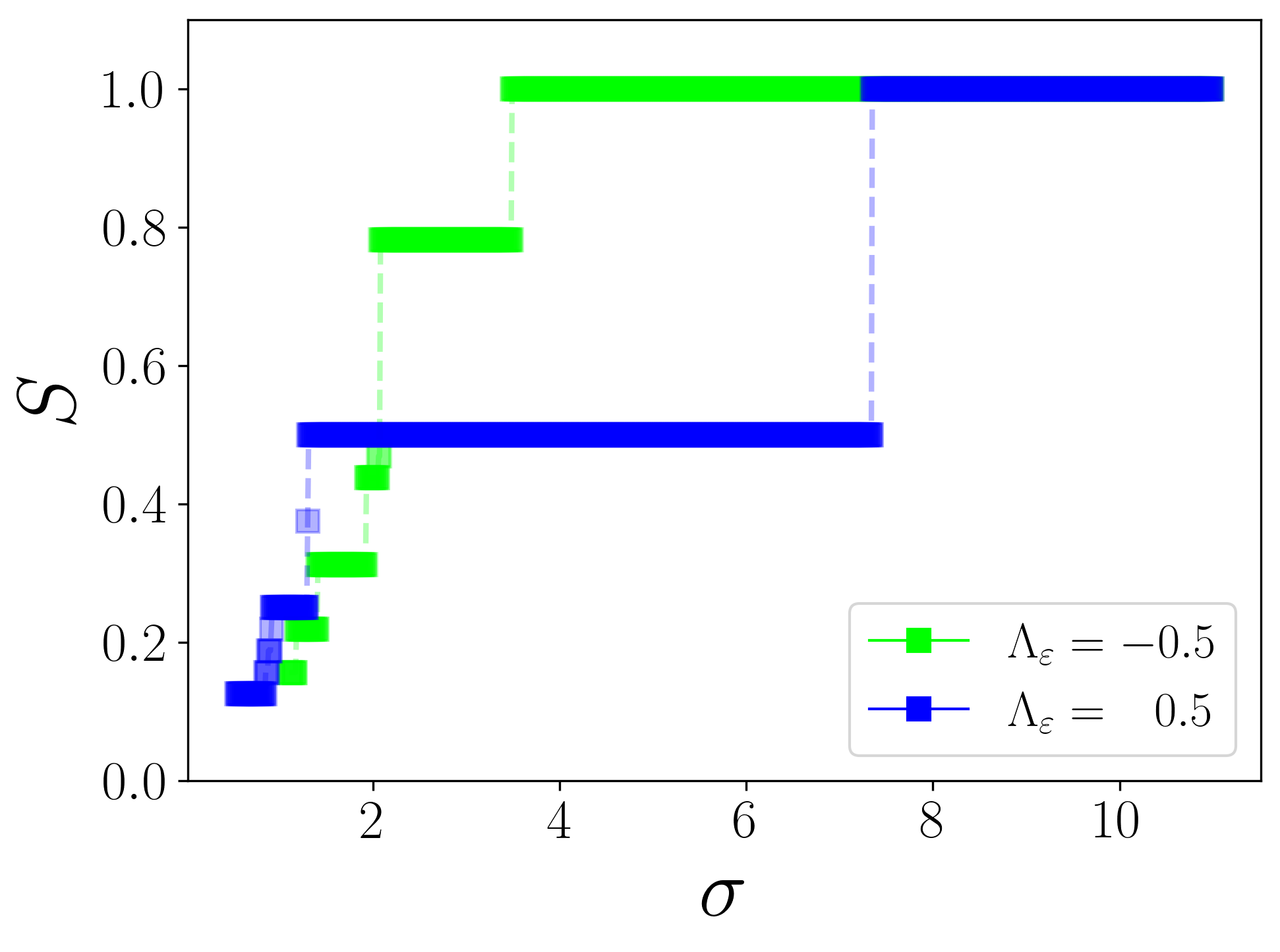}
	\caption{Phase transitions from the evolution equations  for the collective coordinates corresponding to two populations
		each having 4 clusters for $\Lambda_\sigma=0$. Other parameters are $\beta=-0.53\pi$ and  $\varepsilon=0.01$. The natural angular frequencies of clusters ($\Omega_\mu^{\eta}$) are drawn from uniform frequency distribution in the range $[-0.25,0.25]$. See the main text for more details.}
	\label{fig:analytic}
\end{figure}

In the context of mesoscopic description, the synchronization index (\ref{eq:sync_index})  can be redefined to characterize the synchronization of frequency clusters as
\begin{equation}\label{eq:index}
	S = \sum_{\eta,\eta'}\sum_{\mu, \nu} n_\mu^\eta n_\nu^{\eta'} s_{\mu\nu}^{\eta\eta'}.
\end{equation}
Akin to microscopic context (\ref{eq:sync_def}), $s_{\mu \nu}^{\eta\eta'} = 1$, if $\langle \dot{f}_\mu^\eta\rangle = \langle \dot{f}_\nu^{\eta'} \rangle$ otherwise $s_{\mu \nu}^{\eta\eta'} = 0$.
Although the single-step transition can be described by a minimum of two clusters, the multi-step transition requires a larger number of clusters to describe it.
The phase transition displayed by $S$ (\ref{eq:index}), estimated from the evolution equations  for the collective coordinates, corresponding to two populations,
each having 4 clusters, for two  values of  $\Lambda_\varepsilon$ is depicted in Fig.~\ref{fig:analytic}.  It is evident from the figure that the collective coordinates approach
clearly displays single-step and multi-step transitions for $\Lambda_\varepsilon=0.5$ and $-0.5$, respectively, in agreement with the simulation results in Fig.~\ref{fig:transition}.

Now,  an analytical estimate of the upper bound for the coupling strength corresponding to the completely entrained  clusters of intrapopulations (Fig.~\ref{fig:coupling_weights}(g))
during the single-step transition can be obtained  using the perturbative approach in the weak coupling limit~\cite{fial2023}.
Assuming the intracluster phase difference ($f=f_\mu- f_\nu$) grows linearly in time with the relative phase velocity $\Omega^\prime$,  the  collective coordinates
can be expressed in the orders of $\alpha$ as
\begin{subequations}
	\begin{align}
		\Theta_\mu(t)      & = \Theta_\mu^{(0)} + \alpha \Theta_\mu^{(1)}(t) + \mathcal{O}(\alpha^2),           \\
		\kappa_{\mu\nu}(t) & = \kappa_{\mu\nu}^{(0)} + \alpha \kappa_{\mu\nu}^{(1)}(t) + \mathcal{O}(\alpha^2), \\
		f(t)               & = \Omega't +  \alpha f^{(1)}(t) + \mathcal{O}(\alpha^2).
	\end{align}
\end{subequations}

Substituting the perturbed equations in (\ref{eq:collective}) leads to the quadratic equation in  $\Omega^\prime$
\begin{equation}
	(\Omega^\prime)^2 + \Omega^\prime(\Omega_\mu - \Omega_\nu) - \frac{\varepsilon_{\mu\nu} \sigma_{\mu\nu}}{2}\left(r_\mu^{(0)}r_\nu^{(0)}\right)^2 \sin{\beta} = 0.
	\label{eqqe}
\end{equation}
The condition for the existence of real solution of (\ref{eqqe}) requires $ (\Omega_\mu - \Omega_\nu)^2 \geq -2 \varepsilon_{\mu\nu}\sigma_{\mu\nu} (r_\mu^{(0)}r_\nu^{(0)})^2 \sin{\beta}$.
Now, the upper bound for the coupling strength for  the existence of the two-cluster state  can be obtained  as
\begin{equation}
	\sigma_c= \left(\frac{(\Omega_\mu -\Omega_\nu)^2}{2\varepsilon (1-\Lambda_\varepsilon)(1-\Lambda_\sigma) \left(r_\mu^{(0)}r_\nu^{(0)}\right)^2 \sin{(-\beta)}}\right).
	\label{eq:upper}
\end{equation}
It is evident that $\sigma_c$ depends explicitly on the adaptation and the coupling disparity parameters.
For a rough estimate of $ \sigma_c$, one can consider $r_\mu^{(0)} \approx r_\nu^{(0)} \approx 1$.  For the values of the parameters in Fig.~\ref{fig:transition},
the upper bound  can be obtained as $\sigma_c=6.28$ with $\Omega_\mu-\Omega_\nu = -0.25$,
which almost agrees with the single-step transitions in Figs.~\ref{fig:transition} and ~\ref{fig:analytic}.
Now, the critical curve corresponding to the upper bound can be written as $ \Lambda_\varepsilon = 1 - \left[(\Omega_\mu -\Omega_\nu)^2 /\left(2\varepsilon \sigma_c(1-\Lambda_\sigma) (r_\mu^{(0)}r_\nu^{(0)})^2 \sin{(-\beta)} \right) \right]$,  which is depicted as a dashed curve in Figs.~\ref{fig:phase_plots}(a)-\ref{fig:phase_plots}(c) across which the single-step
transition takes place.   The finite size effect and the first order approximation for the order parameters $r_\mu^{(0)}$ and $r_\nu^{(0)}$ contribute to the error between the analytical
curve and the numerical results in Figs.~\ref{fig:phase_plots}.

\textit{Discussion and Conclusion.}
We have considered a  globally coupled finite-size adaptive network, wherein the  subpopulations are distinguished by different degrees of time scale adaptation  $(\Lambda_\varepsilon)$ and
coupling strength $(\Lambda_\sigma)$ but with uniform distribution of natural frequencies.  We have found that nucleations of intrapopulation  frequency clusters
coalesce at the interface of the two populations facilitating the manifestation of  a single large interpopulation frequency cluster for a strong interpopulation
adaptation rate,  $\Lambda_\varepsilon<0$,
without any  coupling strength disparity.   Eventually, the single large inter-frequency-cluster  grows to  the system size as the coupling strength is increased
facilitating a multi-step transition to global synchronization.    In contrast, nucleations of intrapopulation  frequency clusters coalesce among themselves to manifest
completely entrained two intrapopulation frequency clusters as a function of the coupling strength for a strong intrapopulation adaptation rate  $\Lambda_\varepsilon>0$,
without any coupling strength disparity.
Finally, the two intra-frequency-clusters coalesce together facilitating a single-step transition to global synchronization. Synchronization index $S$ clearly displays
the two distinct transitions for two distinct values of $\Lambda_\varepsilon$ when  $\Lambda_\sigma=0$.\\
\indent Further,  we found that a strong interpopulation coupling strength,  $\Lambda_\sigma<0$,  always favors nucleation of inter-frequency-clusters
leading to multi-step transition even for  a strong intrapopulation  adaptation rate  ($\Lambda_\varepsilon>0$).  Furthermore, we found that  a strong intrapopulation
coupling strength,  $\Lambda_\sigma>0$, always favors nucleation  of intra-frequency-clusters facilitating single-step transition even with a  strong interpopulation
adaptation rate  ($\Lambda_\varepsilon<0$). These results corroborate that  the degree of  disparity in the
coupling strength  strongly determines the nature of nucleation leading to distinct synchronization transition.  We   have  analytically  deduced the  macroscopic evolution equations
for the cluster dynamics using the  framework of collective coordinates~\cite{gwga2015}.  The  synchronization transitions obtained using the collective coordinates are found to
agree with  the simulation results.  Further, we  have also deduced the upper bound for the coupling strength for the existence of two intraclusters explicitly in terms of adaptation
and coupling strength disparity parameters, which is found to almost match the  coupling strength at the onset of abrupt single-step transition.
It is also evident that the mesoscopic description brilliantly captures the multicluster dynamics.

Note that similar heterogeneous nucleation resulting in multi- and single-step transitions are reported to be facilitated by distinct quenched disorders~\cite{fial2023}, wherein
the nucleations emerge at the site of the disorder(s)  resulting in multi-(single-)step transitions.  However,  in our case with uniform distribution of natural frequencies,
small intracluster nucleations coalesce  either at the population interface or within the populations resulting in multi- and single-step synchronization transitions depending
on the trade-off between the adaptation and coupling disparities.  Understanding the mechanism of nucleations corresponding to distinct transition due to the inherent disparities
of complex real-world systems is of paramount importance as they shed more light on the role of disparities among different regions of the brain in synchronization  in unraveling
brain functions and neurological disorders~\cite{stam2014},  segregation and polarization dynamics in social networks~\cite{toth2021}, etc. Moreover, our findings hold significance in the
network control theory, offering strategies to optimize adaptive networks.

\textit{Acknowledgments.}
AK acknowledges the financial support from IISER-TVM. JF acknowledges funding by the Austrian Science Fund (FWF): I 5985-N. The work of V.K.C. is supported by DST-CRG Project under Grant No. CRG/2020/004353 and VKC wish to thank DST, New Delhi for computational facilities under the DST-FIST programme (SR/FST/PS- 1/2020/135) to the Department of Physics. DVS  is supported by the DST-SERB-CRG Project under Grant No. CRG/2021/000816.


\begin{thebibliography}{52}%
	\makeatletter
	\providecommand \@ifxundefined [1]{%
	 \@ifx{#1\undefined}
	}%
	\providecommand \@ifnum [1]{%
	 \ifnum #1\expandafter \@firstoftwo
	 \else \expandafter \@secondoftwo
	 \fi
	}%
	\providecommand \@ifx [1]{%
	 \ifx #1\expandafter \@firstoftwo
	 \else \expandafter \@secondoftwo
	 \fi
	}%
	\providecommand \natexlab [1]{#1}%
	\providecommand \enquote  [1]{``#1''}%
	\providecommand \bibnamefont  [1]{#1}%
	\providecommand \bibfnamefont [1]{#1}%
	\providecommand \citenamefont [1]{#1}%
	\providecommand \href@noop [0]{\@secondoftwo}%
	\providecommand \href [0]{\begingroup \@sanitize@url \@href}%
	\providecommand \@href[1]{\@@startlink{#1}\@@href}%
	\providecommand \@@href[1]{\endgroup#1\@@endlink}%
	\providecommand \@sanitize@url [0]{\catcode `\\12\catcode `\$12\catcode `\&12\catcode `\#12\catcode `\^12\catcode `\_12\catcode `\%12\relax}%
	\providecommand \@@startlink[1]{}%
	\providecommand \@@endlink[0]{}%
	\providecommand \url  [0]{\begingroup\@sanitize@url \@url }%
	\providecommand \@url [1]{\endgroup\@href {#1}{\urlprefix }}%
	\providecommand \urlprefix  [0]{URL }%
	\providecommand \Eprint [0]{\href }%
	\providecommand \doibase [0]{https://doi.org/}%
	\providecommand \selectlanguage [0]{\@gobble}%
	\providecommand \bibinfo  [0]{\@secondoftwo}%
	\providecommand \bibfield  [0]{\@secondoftwo}%
	\providecommand \translation [1]{[#1]}%
	\providecommand \BibitemOpen [0]{}%
	\providecommand \bibitemStop [0]{}%
	\providecommand \bibitemNoStop [0]{.\EOS\space}%
	\providecommand \EOS [0]{\spacefactor3000\relax}%
	\providecommand \BibitemShut  [1]{\csname bibitem#1\endcsname}%
	\let\auto@bib@innerbib\@empty
	\bibitem [{\citenamefont {Boccaletti}\ \emph {et~al.}(2006)\citenamefont {Boccaletti}, \citenamefont {Latora}, \citenamefont {Moreno}, \citenamefont {Chavez},\ and\ \citenamefont {Hwang}}]{sbvl2006}%
	  \BibitemOpen
	  \bibfield  {author} {\bibinfo {author} {\bibfnamefont {S.}~\bibnamefont {Boccaletti}}, \bibinfo {author} {\bibfnamefont {V.}~\bibnamefont {Latora}}, \bibinfo {author} {\bibfnamefont {Y.}~\bibnamefont {Moreno}}, \bibinfo {author} {\bibfnamefont {M.}~\bibnamefont {Chavez}},\ and\ \bibinfo {author} {\bibfnamefont {D.-U.}\ \bibnamefont {Hwang}},\ }\bibfield  {title} {\bibinfo {title} {Complex networks: Structure and dynamics},\ }\href {https://doi.org/https://doi.org/10.1016/j.physrep.2005.10.009} {\bibfield  {journal} {\bibinfo  {journal} {Physics Reports}\ }\textbf {\bibinfo {volume} {424}},\ \bibinfo {pages} {175} (\bibinfo {year} {2006})}\BibitemShut {NoStop}%
	\bibitem [{\citenamefont {Strogatz}(2001)}]{strogatz2001}%
	  \BibitemOpen
	  \bibfield  {author} {\bibinfo {author} {\bibfnamefont {S.~H.}\ \bibnamefont {Strogatz}},\ }\bibfield  {title} {\bibinfo {title} {Exploring complex networks},\ }\href@noop {} {\bibfield  {journal} {\bibinfo  {journal} {nature}\ }\textbf {\bibinfo {volume} {410}},\ \bibinfo {pages} {268} (\bibinfo {year} {2001})}\BibitemShut {NoStop}%
	\bibitem [{\citenamefont {Newman}(2003)}]{newman2003}%
	  \BibitemOpen
	  \bibfield  {author} {\bibinfo {author} {\bibfnamefont {M.~E.~J.}\ \bibnamefont {Newman}},\ }\bibfield  {title} {\bibinfo {title} {The structure and function of complex networks},\ }\href {https://doi.org/10.1137/S003614450342480} {\bibfield  {journal} {\bibinfo  {journal} {SIAM Review}\ }\textbf {\bibinfo {volume} {45}},\ \bibinfo {pages} {167} (\bibinfo {year} {2003})}\BibitemShut {NoStop}%
	\bibitem [{\citenamefont {Boccaletti}\ \emph {et~al.}(2018)\citenamefont {Boccaletti}, \citenamefont {Pisarchik}, \citenamefont {Del~Genio},\ and\ \citenamefont {Amann}}]{boccaletti2018}%
	  \BibitemOpen
	  \bibfield  {author} {\bibinfo {author} {\bibfnamefont {S.}~\bibnamefont {Boccaletti}}, \bibinfo {author} {\bibfnamefont {A.~N.}\ \bibnamefont {Pisarchik}}, \bibinfo {author} {\bibfnamefont {C.~I.}\ \bibnamefont {Del~Genio}},\ and\ \bibinfo {author} {\bibfnamefont {A.}~\bibnamefont {Amann}},\ }\href@noop {} {\emph {\bibinfo {title} {Synchronization: from coupled systems to complex networks}}}\ (\bibinfo  {publisher} {Cambridge University Press},\ \bibinfo {year} {2018})\BibitemShut {NoStop}%
	\bibitem [{\citenamefont {D{\"o}rfler}\ \emph {et~al.}(2013)\citenamefont {D{\"o}rfler}, \citenamefont {Chertkov},\ and\ \citenamefont {Bullo}}]{dorfler2013}%
	  \BibitemOpen
	  \bibfield  {author} {\bibinfo {author} {\bibfnamefont {F.}~\bibnamefont {D{\"o}rfler}}, \bibinfo {author} {\bibfnamefont {M.}~\bibnamefont {Chertkov}},\ and\ \bibinfo {author} {\bibfnamefont {F.}~\bibnamefont {Bullo}},\ }\bibfield  {title} {\bibinfo {title} {Synchronization in complex oscillator networks and smart grids},\ }\href@noop {} {\bibfield  {journal} {\bibinfo  {journal} {Proceedings of the National Academy of Sciences}\ }\textbf {\bibinfo {volume} {110}},\ \bibinfo {pages} {2005} (\bibinfo {year} {2013})}\BibitemShut {NoStop}%
	\bibitem [{\citenamefont {Shahal}\ \emph {et~al.}(2020)\citenamefont {Shahal}, \citenamefont {Wurzberg}, \citenamefont {Sibony}, \citenamefont {Duadi}, \citenamefont {Shniderman}, \citenamefont {Weymouth}, \citenamefont {Davidson},\ and\ \citenamefont {Fridman}}]{shahal2020}%
	  \BibitemOpen
	  \bibfield  {author} {\bibinfo {author} {\bibfnamefont {S.}~\bibnamefont {Shahal}}, \bibinfo {author} {\bibfnamefont {A.}~\bibnamefont {Wurzberg}}, \bibinfo {author} {\bibfnamefont {I.}~\bibnamefont {Sibony}}, \bibinfo {author} {\bibfnamefont {H.}~\bibnamefont {Duadi}}, \bibinfo {author} {\bibfnamefont {E.}~\bibnamefont {Shniderman}}, \bibinfo {author} {\bibfnamefont {D.}~\bibnamefont {Weymouth}}, \bibinfo {author} {\bibfnamefont {N.}~\bibnamefont {Davidson}},\ and\ \bibinfo {author} {\bibfnamefont {M.}~\bibnamefont {Fridman}},\ }\bibfield  {title} {\bibinfo {title} {Synchronization of complex human networks},\ }\href@noop {} {\bibfield  {journal} {\bibinfo  {journal} {Nature communications}\ }\textbf {\bibinfo {volume} {11}},\ \bibinfo {pages} {3854} (\bibinfo {year} {2020})}\BibitemShut {NoStop}%
	\bibitem [{\citenamefont {Bagchi}\ and\ \citenamefont {Mohanty}(2011)}]{bdmp2011}%
	  \BibitemOpen
	  \bibfield  {author} {\bibinfo {author} {\bibfnamefont {D.}~\bibnamefont {Bagchi}}\ and\ \bibinfo {author} {\bibfnamefont {P.~K.}\ \bibnamefont {Mohanty}},\ }\bibfield  {title} {\bibinfo {title} {Phase transition in an exactly solvable extinction model},\ }\href {https://doi.org/10.1103/PhysRevE.84.061921} {\bibfield  {journal} {\bibinfo  {journal} {Phys. Rev. E}\ }\textbf {\bibinfo {volume} {84}},\ \bibinfo {pages} {061921} (\bibinfo {year} {2011})}\BibitemShut {NoStop}%
	\bibitem [{\citenamefont {Conover}\ \emph {et~al.}(2021)\citenamefont {Conover}, \citenamefont {Ratkiewicz}, \citenamefont {Francisco}, \citenamefont {Goncalves}, \citenamefont {Menczer},\ and\ \citenamefont {Flammini}}]{conover2011}%
	  \BibitemOpen
	  \bibfield  {author} {\bibinfo {author} {\bibfnamefont {M.}~\bibnamefont {Conover}}, \bibinfo {author} {\bibfnamefont {J.}~\bibnamefont {Ratkiewicz}}, \bibinfo {author} {\bibfnamefont {M.}~\bibnamefont {Francisco}}, \bibinfo {author} {\bibfnamefont {B.}~\bibnamefont {Goncalves}}, \bibinfo {author} {\bibfnamefont {F.}~\bibnamefont {Menczer}},\ and\ \bibinfo {author} {\bibfnamefont {A.}~\bibnamefont {Flammini}},\ }\bibfield  {title} {\bibinfo {title} {Political polarization on twitter},\ }\href {https://doi.org/10.1609/icwsm.v5i1.14126} {\bibfield  {journal} {\bibinfo  {journal} {Proceedings of the International AAAI Conference on Web and Social Media}\ }\textbf {\bibinfo {volume} {5}},\ \bibinfo {pages} {89} (\bibinfo {year} {2021})}\BibitemShut {NoStop}%
	\bibitem [{\citenamefont {Levy}(2008)}]{levy2008stock}%
	  \BibitemOpen
	  \bibfield  {author} {\bibinfo {author} {\bibfnamefont {M.}~\bibnamefont {Levy}},\ }\bibfield  {title} {\bibinfo {title} {Stock market crashes as social phase transitions},\ }\href@noop {} {\bibfield  {journal} {\bibinfo  {journal} {Journal of Economic Dynamics and Control}\ }\textbf {\bibinfo {volume} {32}},\ \bibinfo {pages} {137} (\bibinfo {year} {2008})}\BibitemShut {NoStop}%
	\bibitem [{\citenamefont {Arenas}\ \emph {et~al.}(2008)\citenamefont {Arenas}, \citenamefont {Díaz-Guilera}, \citenamefont {Kurths}, \citenamefont {Moreno},\ and\ \citenamefont {Zhou}}]{alex2008}%
	  \BibitemOpen
	  \bibfield  {author} {\bibinfo {author} {\bibfnamefont {A.}~\bibnamefont {Arenas}}, \bibinfo {author} {\bibfnamefont {A.}~\bibnamefont {Díaz-Guilera}}, \bibinfo {author} {\bibfnamefont {J.}~\bibnamefont {Kurths}}, \bibinfo {author} {\bibfnamefont {Y.}~\bibnamefont {Moreno}},\ and\ \bibinfo {author} {\bibfnamefont {C.}~\bibnamefont {Zhou}},\ }\bibfield  {title} {\bibinfo {title} {Synchronization in complex networks},\ }\href {https://doi.org/https://doi.org/10.1016/j.physrep.2008.09.002} {\bibfield  {journal} {\bibinfo  {journal} {Physics Reports}\ }\textbf {\bibinfo {volume} {469}},\ \bibinfo {pages} {93} (\bibinfo {year} {2008})}\BibitemShut {NoStop}%
	\bibitem [{\citenamefont {G\'omez-Garde\~nes}\ \emph {et~al.}(2007)\citenamefont {G\'omez-Garde\~nes}, \citenamefont {Moreno},\ and\ \citenamefont {Arenas}}]{gjya2007}%
	  \BibitemOpen
	  \bibfield  {author} {\bibinfo {author} {\bibfnamefont {J.}~\bibnamefont {G\'omez-Garde\~nes}}, \bibinfo {author} {\bibfnamefont {Y.}~\bibnamefont {Moreno}},\ and\ \bibinfo {author} {\bibfnamefont {A.}~\bibnamefont {Arenas}},\ }\bibfield  {title} {\bibinfo {title} {Paths to synchronization on complex networks},\ }\href {https://doi.org/10.1103/PhysRevLett.98.034101} {\bibfield  {journal} {\bibinfo  {journal} {Phys. Rev. Lett.}\ }\textbf {\bibinfo {volume} {98}},\ \bibinfo {pages} {034101} (\bibinfo {year} {2007})}\BibitemShut {NoStop}%
	\bibitem [{\citenamefont {Dutta}\ \emph {et~al.}(2023)\citenamefont {Dutta}, \citenamefont {Kundu}, \citenamefont {Khanra}, \citenamefont {Hens},\ and\ \citenamefont {Pal}}]{dskp2023}%
	  \BibitemOpen
	  \bibfield  {author} {\bibinfo {author} {\bibfnamefont {S.}~\bibnamefont {Dutta}}, \bibinfo {author} {\bibfnamefont {P.}~\bibnamefont {Kundu}}, \bibinfo {author} {\bibfnamefont {P.}~\bibnamefont {Khanra}}, \bibinfo {author} {\bibfnamefont {C.}~\bibnamefont {Hens}},\ and\ \bibinfo {author} {\bibfnamefont {P.}~\bibnamefont {Pal}},\ }\bibfield  {title} {\bibinfo {title} {Perfect synchronization in complex networks with higher-order interactions},\ }\href {https://doi.org/10.1103/PhysRevE.108.024304} {\bibfield  {journal} {\bibinfo  {journal} {Phys. Rev. E}\ }\textbf {\bibinfo {volume} {108}},\ \bibinfo {pages} {024304} (\bibinfo {year} {2023})}\BibitemShut {NoStop}%
	\bibitem [{\citenamefont {Gerstner}\ \emph {et~al.}(1996)\citenamefont {Gerstner}, \citenamefont {Kempter}, \citenamefont {Van~Hemmen},\ and\ \citenamefont {Wagner}}]{gerstner1996}%
	  \BibitemOpen
	  \bibfield  {author} {\bibinfo {author} {\bibfnamefont {W.}~\bibnamefont {Gerstner}}, \bibinfo {author} {\bibfnamefont {R.}~\bibnamefont {Kempter}}, \bibinfo {author} {\bibfnamefont {J.~L.}\ \bibnamefont {Van~Hemmen}},\ and\ \bibinfo {author} {\bibfnamefont {H.}~\bibnamefont {Wagner}},\ }\bibfield  {title} {\bibinfo {title} {A neuronal learning rule for sub-millisecond temporal coding},\ }\href@noop {} {\bibfield  {journal} {\bibinfo  {journal} {Nature}\ }\textbf {\bibinfo {volume} {383}},\ \bibinfo {pages} {76} (\bibinfo {year} {1996})}\BibitemShut {NoStop}%
	\bibitem [{\citenamefont {Caporale}\ and\ \citenamefont {Dan}(2008)}]{cndy2008}%
	  \BibitemOpen
	  \bibfield  {author} {\bibinfo {author} {\bibfnamefont {N.}~\bibnamefont {Caporale}}\ and\ \bibinfo {author} {\bibfnamefont {Y.}~\bibnamefont {Dan}},\ }\bibfield  {title} {\bibinfo {title} {Spike timing-dependent plasticity: A hebbian learning rule},\ }\href {https://doi.org/10.1146/annurev.neuro.31.060407.125639} {\bibfield  {journal} {\bibinfo  {journal} {Annual Review of Neuroscience}\ }\textbf {\bibinfo {volume} {31}},\ \bibinfo {pages} {25} (\bibinfo {year} {2008})},\ \bibinfo {note} {pMID: 18275283}\BibitemShut {NoStop}%
	\bibitem [{\citenamefont {R{\"o}hr}\ \emph {et~al.}(2019)\citenamefont {R{\"o}hr}, \citenamefont {Berner}, \citenamefont {Lameu}, \citenamefont {Popovych},\ and\ \citenamefont {Yanchuk}}]{rohr2019}%
	  \BibitemOpen
	  \bibfield  {author} {\bibinfo {author} {\bibfnamefont {V.}~\bibnamefont {R{\"o}hr}}, \bibinfo {author} {\bibfnamefont {R.}~\bibnamefont {Berner}}, \bibinfo {author} {\bibfnamefont {E.~L.}\ \bibnamefont {Lameu}}, \bibinfo {author} {\bibfnamefont {O.~V.}\ \bibnamefont {Popovych}},\ and\ \bibinfo {author} {\bibfnamefont {S.}~\bibnamefont {Yanchuk}},\ }\bibfield  {title} {\bibinfo {title} {Frequency cluster formation and slow oscillations in neural populations with plasticity},\ }\href@noop {} {\bibfield  {journal} {\bibinfo  {journal} {PLoS One}\ }\textbf {\bibinfo {volume} {14}},\ \bibinfo {pages} {e0225094} (\bibinfo {year} {2019})}\BibitemShut {NoStop}%
	\bibitem [{\citenamefont {Waldrop}(2013)}]{waldrop2013}%
	  \BibitemOpen
	  \bibfield  {author} {\bibinfo {author} {\bibfnamefont {M.~M.}\ \bibnamefont {Waldrop}},\ }\bibfield  {title} {\bibinfo {title} {Smart connections},\ }\href@noop {} {\bibfield  {journal} {\bibinfo  {journal} {Nature}\ }\textbf {\bibinfo {volume} {503}},\ \bibinfo {pages} {22} (\bibinfo {year} {2013})}\BibitemShut {NoStop}%
	\bibitem [{\citenamefont {Morales}\ \emph {et~al.}(2021)\citenamefont {Morales}, \citenamefont {Mirasso},\ and\ \citenamefont {Soriano}}]{MORALES2021}%
	  \BibitemOpen
	  \bibfield  {author} {\bibinfo {author} {\bibfnamefont {G.~B.}\ \bibnamefont {Morales}}, \bibinfo {author} {\bibfnamefont {C.~R.}\ \bibnamefont {Mirasso}},\ and\ \bibinfo {author} {\bibfnamefont {M.~C.}\ \bibnamefont {Soriano}},\ }\bibfield  {title} {\bibinfo {title} {Unveiling the role of plasticity rules in reservoir computing},\ }\href {https://doi.org/https://doi.org/10.1016/j.neucom.2020.05.127} {\bibfield  {journal} {\bibinfo  {journal} {Neurocomputing}\ }\textbf {\bibinfo {volume} {461}},\ \bibinfo {pages} {705} (\bibinfo {year} {2021})}\BibitemShut {NoStop}%
	\bibitem [{\citenamefont {Berner}\ \emph {et~al.}(2021)\citenamefont {Berner}, \citenamefont {Yanchuk},\ and\ \citenamefont {Sch\"oll}}]{brys2021}%
	  \BibitemOpen
	  \bibfield  {author} {\bibinfo {author} {\bibfnamefont {R.}~\bibnamefont {Berner}}, \bibinfo {author} {\bibfnamefont {S.}~\bibnamefont {Yanchuk}},\ and\ \bibinfo {author} {\bibfnamefont {E.}~\bibnamefont {Sch\"oll}},\ }\bibfield  {title} {\bibinfo {title} {What adaptive neuronal networks teach us about power grids},\ }\href {https://doi.org/10.1103/PhysRevE.103.042315} {\bibfield  {journal} {\bibinfo  {journal} {Phys. Rev. E}\ }\textbf {\bibinfo {volume} {103}},\ \bibinfo {pages} {042315} (\bibinfo {year} {2021})}\BibitemShut {NoStop}%
	\bibitem [{\citenamefont {Proulx}\ \emph {et~al.}(2005)\citenamefont {Proulx}, \citenamefont {Promislow},\ and\ \citenamefont {Phillips}}]{PROULX2005}%
	  \BibitemOpen
	  \bibfield  {author} {\bibinfo {author} {\bibfnamefont {S.~R.}\ \bibnamefont {Proulx}}, \bibinfo {author} {\bibfnamefont {D.~E.}\ \bibnamefont {Promislow}},\ and\ \bibinfo {author} {\bibfnamefont {P.~C.}\ \bibnamefont {Phillips}},\ }\bibfield  {title} {\bibinfo {title} {Network thinking in ecology and evolution},\ }\href {https://doi.org/https://doi.org/10.1016/j.tree.2005.04.004} {\bibfield  {journal} {\bibinfo  {journal} {Trends in Ecology \& Evolution}\ }\textbf {\bibinfo {volume} {20}},\ \bibinfo {pages} {345} (\bibinfo {year} {2005})}\BibitemShut {NoStop}%
	\bibitem [{\citenamefont {Gross}\ \emph {et~al.}(2006)\citenamefont {Gross}, \citenamefont {D'Lima},\ and\ \citenamefont {Blasius}}]{gtdc2006}%
	  \BibitemOpen
	  \bibfield  {author} {\bibinfo {author} {\bibfnamefont {T.}~\bibnamefont {Gross}}, \bibinfo {author} {\bibfnamefont {C.~J.~D.}\ \bibnamefont {D'Lima}},\ and\ \bibinfo {author} {\bibfnamefont {B.}~\bibnamefont {Blasius}},\ }\bibfield  {title} {\bibinfo {title} {Epidemic dynamics on an adaptive network},\ }\href {https://doi.org/10.1103/PhysRevLett.96.208701} {\bibfield  {journal} {\bibinfo  {journal} {Phys. Rev. Lett.}\ }\textbf {\bibinfo {volume} {96}},\ \bibinfo {pages} {208701} (\bibinfo {year} {2006})}\BibitemShut {NoStop}%
	\bibitem [{\citenamefont {Rajapakse}\ \emph {et~al.}(2011)\citenamefont {Rajapakse}, \citenamefont {Groudine},\ and\ \citenamefont {Mesbahi}}]{irmg2011}%
	  \BibitemOpen
	  \bibfield  {author} {\bibinfo {author} {\bibfnamefont {I.}~\bibnamefont {Rajapakse}}, \bibinfo {author} {\bibfnamefont {M.}~\bibnamefont {Groudine}},\ and\ \bibinfo {author} {\bibfnamefont {M.}~\bibnamefont {Mesbahi}},\ }\bibfield  {title} {\bibinfo {title} {Dynamics and control of state-dependent networks for probing genomic organization},\ }\href {https://doi.org/10.1073/pnas.1113249108} {\bibfield  {journal} {\bibinfo  {journal} {Proceedings of the National Academy of Sciences}\ }\textbf {\bibinfo {volume} {108}},\ \bibinfo {pages} {17257} (\bibinfo {year} {2011})}\BibitemShut {NoStop}%
	\bibitem [{\citenamefont {Horstmeyer}\ and\ \citenamefont {Kuehn}(2020)}]{hlkc2020}%
	  \BibitemOpen
	  \bibfield  {author} {\bibinfo {author} {\bibfnamefont {L.}~\bibnamefont {Horstmeyer}}\ and\ \bibinfo {author} {\bibfnamefont {C.}~\bibnamefont {Kuehn}},\ }\bibfield  {title} {\bibinfo {title} {Adaptive voter model on simplicial complexes},\ }\href {https://doi.org/10.1103/PhysRevE.101.022305} {\bibfield  {journal} {\bibinfo  {journal} {Phys. Rev. E}\ }\textbf {\bibinfo {volume} {101}},\ \bibinfo {pages} {022305} (\bibinfo {year} {2020})}\BibitemShut {NoStop}%
	\bibitem [{\citenamefont {Baumann}\ \emph {et~al.}(2020)\citenamefont {Baumann}, \citenamefont {Lorenz-Spreen}, \citenamefont {Sokolov},\ and\ \citenamefont {Starnini}}]{bfls2020}%
	  \BibitemOpen
	  \bibfield  {author} {\bibinfo {author} {\bibfnamefont {F.}~\bibnamefont {Baumann}}, \bibinfo {author} {\bibfnamefont {P.}~\bibnamefont {Lorenz-Spreen}}, \bibinfo {author} {\bibfnamefont {I.~M.}\ \bibnamefont {Sokolov}},\ and\ \bibinfo {author} {\bibfnamefont {M.}~\bibnamefont {Starnini}},\ }\bibfield  {title} {\bibinfo {title} {Modeling echo chambers and polarization dynamics in social networks},\ }\href {https://doi.org/10.1103/PhysRevLett.124.048301} {\bibfield  {journal} {\bibinfo  {journal} {Phys. Rev. Lett.}\ }\textbf {\bibinfo {volume} {124}},\ \bibinfo {pages} {048301} (\bibinfo {year} {2020})}\BibitemShut {NoStop}%
	\bibitem [{\citenamefont {Antoniades}\ and\ \citenamefont {Dovrolis}(2015)}]{addc2015}%
	  \BibitemOpen
	  \bibfield  {author} {\bibinfo {author} {\bibfnamefont {D.}~\bibnamefont {Antoniades}}\ and\ \bibinfo {author} {\bibfnamefont {C.}~\bibnamefont {Dovrolis}},\ }\bibfield  {title} {\bibinfo {title} {Co-evolutionary dynamics in social networks: A case study of twitter},\ }\href@noop {} {\bibfield  {journal} {\bibinfo  {journal} {Computational Social Networks}\ }\textbf {\bibinfo {volume} {2}},\ \bibinfo {pages} {1} (\bibinfo {year} {2015})}\BibitemShut {NoStop}%
	\bibitem [{\citenamefont {Sawicki}\ \emph {et~al.}(2023)\citenamefont {Sawicki}, \citenamefont {Berner}, \citenamefont {Loos}, \citenamefont {Anvari}, \citenamefont {Bader}, \citenamefont {Barfuss}, \citenamefont {Botta}, \citenamefont {Brede}, \citenamefont {Franović}, \citenamefont {Gauthier}, \citenamefont {Goldt}, \citenamefont {Hajizadeh}, \citenamefont {Hövel}, \citenamefont {Karin}, \citenamefont {Lorenz-Spreen}, \citenamefont {Miehl}, \citenamefont {Mölter}, \citenamefont {Olmi}, \citenamefont {Schöll}, \citenamefont {Seif}, \citenamefont {Tass}, \citenamefont {Volpe}, \citenamefont {Yanchuk},\ and\ \citenamefont {Kurths}}]{sjbr2023}%
	  \BibitemOpen
	  \bibfield  {author} {\bibinfo {author} {\bibfnamefont {J.}~\bibnamefont {Sawicki}}, \bibinfo {author} {\bibfnamefont {R.}~\bibnamefont {Berner}}, \bibinfo {author} {\bibfnamefont {S.~A.~M.}\ \bibnamefont {Loos}}, \bibinfo {author} {\bibfnamefont {M.}~\bibnamefont {Anvari}}, \bibinfo {author} {\bibfnamefont {R.}~\bibnamefont {Bader}}, \bibinfo {author} {\bibfnamefont {W.}~\bibnamefont {Barfuss}}, \bibinfo {author} {\bibfnamefont {N.}~\bibnamefont {Botta}}, \bibinfo {author} {\bibfnamefont {N.}~\bibnamefont {Brede}}, \bibinfo {author} {\bibfnamefont {I.}~\bibnamefont {Franović}}, \bibinfo {author} {\bibfnamefont {D.~J.}\ \bibnamefont {Gauthier}}, \bibinfo {author} {\bibfnamefont {S.}~\bibnamefont {Goldt}}, \bibinfo {author} {\bibfnamefont {A.}~\bibnamefont {Hajizadeh}}, \bibinfo {author} {\bibfnamefont {P.}~\bibnamefont {Hövel}}, \bibinfo {author} {\bibfnamefont {O.}~\bibnamefont {Karin}}, \bibinfo {author} {\bibfnamefont {P.}~\bibnamefont {Lorenz-Spreen}}, \bibinfo {author} {\bibfnamefont {C.}~\bibnamefont {Miehl}}, \bibinfo {author} {\bibfnamefont {J.}~\bibnamefont {Mölter}}, \bibinfo {author} {\bibfnamefont {S.}~\bibnamefont {Olmi}}, \bibinfo {author} {\bibfnamefont {E.}~\bibnamefont {Schöll}}, \bibinfo {author} {\bibfnamefont {A.}~\bibnamefont {Seif}}, \bibinfo {author} {\bibfnamefont {P.~A.}\ \bibnamefont {Tass}}, \bibinfo {author} {\bibfnamefont {G.}~\bibnamefont {Volpe}}, \bibinfo {author} {\bibfnamefont {S.}~\bibnamefont {Yanchuk}},\ and\ \bibinfo {author} {\bibfnamefont {J.}~\bibnamefont {Kurths}},\ }\bibfield  {title} {\bibinfo {title} {Perspectives on adaptive dynamical systems},\ }\href {https://doi.org/10.1063/5.0147231} {\bibfield  {journal} {\bibinfo  {journal} {Chaos: An Interdisciplinary Journal of Nonlinear Science}\ }\textbf {\bibinfo {volume} {33}},\ \bibinfo {pages} {071501} (\bibinfo {year} {2023})}\BibitemShut {NoStop}%
	\bibitem [{\citenamefont {Jüttner}\ and\ \citenamefont {Martens}(2023)}]{jube2023}%
	  \BibitemOpen
	  \bibfield  {author} {\bibinfo {author} {\bibfnamefont {B.}~\bibnamefont {Jüttner}}\ and\ \bibinfo {author} {\bibfnamefont {E.~A.}\ \bibnamefont {Martens}},\ }\bibfield  {title} {\bibinfo {title} {Complex dynamics in adaptive phase oscillator networks},\ }\href {https://doi.org/10.1063/5.0133190} {\bibfield  {journal} {\bibinfo  {journal} {Chaos: An Interdisciplinary Journal of Nonlinear Science}\ }\textbf {\bibinfo {volume} {33}},\ \bibinfo {pages} {053106} (\bibinfo {year} {2023})}\BibitemShut {NoStop}%
	\bibitem [{\citenamefont {Berner}(2021)}]{berner2021}%
	  \BibitemOpen
	  \bibfield  {author} {\bibinfo {author} {\bibfnamefont {R.}~\bibnamefont {Berner}},\ }\href {https://doi.org/10.1007/978-3-030-74938-5} {\emph {\bibinfo {title} {Patterns of synchrony in complex networks of adaptively coupled oscillators}}}\ (\bibinfo  {publisher} {Springer Theses},\ \bibinfo {year} {2021})\BibitemShut {NoStop}%
	\bibitem [{\citenamefont {Berner}\ \emph {et~al.}(2019)\citenamefont {Berner}, \citenamefont {Scholl},\ and\ \citenamefont {Yanchuk}}]{berner2019multi}%
	  \BibitemOpen
	  \bibfield  {author} {\bibinfo {author} {\bibfnamefont {R.}~\bibnamefont {Berner}}, \bibinfo {author} {\bibfnamefont {E.}~\bibnamefont {Scholl}},\ and\ \bibinfo {author} {\bibfnamefont {S.}~\bibnamefont {Yanchuk}},\ }\bibfield  {title} {\bibinfo {title} {Multiclusters in networks of adaptively coupled phase oscillators},\ }\href@noop {} {\bibfield  {journal} {\bibinfo  {journal} {SIAM Journal on Applied Dynamical Systems}\ }\textbf {\bibinfo {volume} {18}},\ \bibinfo {pages} {2227} (\bibinfo {year} {2019})}\BibitemShut {NoStop}%
	\bibitem [{\citenamefont {Thamizharasan}\ \emph {et~al.}(2022)\citenamefont {Thamizharasan}, \citenamefont {Chandrasekar}, \citenamefont {Senthilvelan}, \citenamefont {Berner}, \citenamefont {Sch\"oll},\ and\ \citenamefont {Senthilkumar}}]{thch2022}%
	  \BibitemOpen
	  \bibfield  {author} {\bibinfo {author} {\bibfnamefont {S.}~\bibnamefont {Thamizharasan}}, \bibinfo {author} {\bibfnamefont {V.~K.}\ \bibnamefont {Chandrasekar}}, \bibinfo {author} {\bibfnamefont {M.}~\bibnamefont {Senthilvelan}}, \bibinfo {author} {\bibfnamefont {R.}~\bibnamefont {Berner}}, \bibinfo {author} {\bibfnamefont {E.}~\bibnamefont {Sch\"oll}},\ and\ \bibinfo {author} {\bibfnamefont {D.~V.}\ \bibnamefont {Senthilkumar}},\ }\bibfield  {title} {\bibinfo {title} {Exotic states induced by coevolving connection weights and phases in complex networks},\ }\href {https://doi.org/10.1103/PhysRevE.105.034312} {\bibfield  {journal} {\bibinfo  {journal} {Phys. Rev. E}\ }\textbf {\bibinfo {volume} {105}},\ \bibinfo {pages} {034312} (\bibinfo {year} {2022})}\BibitemShut {NoStop}%
	\bibitem [{\citenamefont {Berner}\ \emph {et~al.}(2020)\citenamefont {Berner}, \citenamefont {Polanska}, \citenamefont {Sch{\"o}ll},\ and\ \citenamefont {Yanchuk}}]{berner2020solitary}%
	  \BibitemOpen
	  \bibfield  {author} {\bibinfo {author} {\bibfnamefont {R.}~\bibnamefont {Berner}}, \bibinfo {author} {\bibfnamefont {A.}~\bibnamefont {Polanska}}, \bibinfo {author} {\bibfnamefont {E.}~\bibnamefont {Sch{\"o}ll}},\ and\ \bibinfo {author} {\bibfnamefont {S.}~\bibnamefont {Yanchuk}},\ }\bibfield  {title} {\bibinfo {title} {Solitary states in adaptive nonlocal oscillator networks},\ }\href@noop {} {\bibfield  {journal} {\bibinfo  {journal} {The European Physical Journal Special Topics}\ }\textbf {\bibinfo {volume} {229}},\ \bibinfo {pages} {2183} (\bibinfo {year} {2020})}\BibitemShut {NoStop}%
	\bibitem [{\citenamefont {Thiele}\ \emph {et~al.}(2023)\citenamefont {Thiele}, \citenamefont {Berner}, \citenamefont {Tass}, \citenamefont {Sch{\"o}ll},\ and\ \citenamefont {Yanchuk}}]{thiele2023asymmetric}%
	  \BibitemOpen
	  \bibfield  {author} {\bibinfo {author} {\bibfnamefont {M.}~\bibnamefont {Thiele}}, \bibinfo {author} {\bibfnamefont {R.}~\bibnamefont {Berner}}, \bibinfo {author} {\bibfnamefont {P.~A.}\ \bibnamefont {Tass}}, \bibinfo {author} {\bibfnamefont {E.}~\bibnamefont {Sch{\"o}ll}},\ and\ \bibinfo {author} {\bibfnamefont {S.}~\bibnamefont {Yanchuk}},\ }\bibfield  {title} {\bibinfo {title} {Asymmetric adaptivity induces recurrent synchronization in complex networks},\ }\href@noop {} {\bibfield  {journal} {\bibinfo  {journal} {Chaos: An Interdisciplinary Journal of Nonlinear Science}\ }\textbf {\bibinfo {volume} {33}} (\bibinfo {year} {2023})}\BibitemShut {NoStop}%
	\bibitem [{\citenamefont {Fialkowski}\ \emph {et~al.}(2023)\citenamefont {Fialkowski}, \citenamefont {Yanchuk}, \citenamefont {Sokolov}, \citenamefont {Sch\"oll}, \citenamefont {Gottwald},\ and\ \citenamefont {Berner}}]{fial2023}%
	  \BibitemOpen
	  \bibfield  {author} {\bibinfo {author} {\bibfnamefont {J.}~\bibnamefont {Fialkowski}}, \bibinfo {author} {\bibfnamefont {S.}~\bibnamefont {Yanchuk}}, \bibinfo {author} {\bibfnamefont {I.~M.}\ \bibnamefont {Sokolov}}, \bibinfo {author} {\bibfnamefont {E.}~\bibnamefont {Sch\"oll}}, \bibinfo {author} {\bibfnamefont {G.~A.}\ \bibnamefont {Gottwald}},\ and\ \bibinfo {author} {\bibfnamefont {R.}~\bibnamefont {Berner}},\ }\bibfield  {title} {\bibinfo {title} {Heterogeneous nucleation in finite-size adaptive dynamical networks},\ }\href {https://doi.org/10.1103/PhysRevLett.130.067402} {\bibfield  {journal} {\bibinfo  {journal} {Phys. Rev. Lett.}\ }\textbf {\bibinfo {volume} {130}},\ \bibinfo {pages} {067402} (\bibinfo {year} {2023})}\BibitemShut {NoStop}%
	\bibitem [{\citenamefont {Berner}\ \emph {et~al.}(2023)\citenamefont {Berner}, \citenamefont {Gross}, \citenamefont {Kuehn}, \citenamefont {Kurths},\ and\ \citenamefont {Yanchuk}}]{BERNER2023}%
	  \BibitemOpen
	  \bibfield  {author} {\bibinfo {author} {\bibfnamefont {R.}~\bibnamefont {Berner}}, \bibinfo {author} {\bibfnamefont {T.}~\bibnamefont {Gross}}, \bibinfo {author} {\bibfnamefont {C.}~\bibnamefont {Kuehn}}, \bibinfo {author} {\bibfnamefont {J.}~\bibnamefont {Kurths}},\ and\ \bibinfo {author} {\bibfnamefont {S.}~\bibnamefont {Yanchuk}},\ }\bibfield  {title} {\bibinfo {title} {Adaptive dynamical networks},\ }\href {https://doi.org/https://doi.org/10.1016/j.physrep.2023.08.001} {\bibfield  {journal} {\bibinfo  {journal} {Physics Reports}\ }\textbf {\bibinfo {volume} {1031}},\ \bibinfo {pages} {1} (\bibinfo {year} {2023})},\ \bibinfo {note} {adaptive dynamical networks}\BibitemShut {NoStop}%
	\bibitem [{\citenamefont {Girvan}\ and\ \citenamefont {Newman}(2002)}]{girvan2002}%
	  \BibitemOpen
	  \bibfield  {author} {\bibinfo {author} {\bibfnamefont {M.}~\bibnamefont {Girvan}}\ and\ \bibinfo {author} {\bibfnamefont {M.~E.}\ \bibnamefont {Newman}},\ }\bibfield  {title} {\bibinfo {title} {Community structure in social and biological networks},\ }\href@noop {} {\bibfield  {journal} {\bibinfo  {journal} {Proceedings of the national academy of sciences}\ }\textbf {\bibinfo {volume} {99}},\ \bibinfo {pages} {7821} (\bibinfo {year} {2002})}\BibitemShut {NoStop}%
	\bibitem [{\citenamefont {Newman}(2006)}]{newman2006}%
	  \BibitemOpen
	  \bibfield  {author} {\bibinfo {author} {\bibfnamefont {M.~E.}\ \bibnamefont {Newman}},\ }\bibfield  {title} {\bibinfo {title} {Modularity and community structure in networks},\ }\href@noop {} {\bibfield  {journal} {\bibinfo  {journal} {Proceedings of the national academy of sciences}\ }\textbf {\bibinfo {volume} {103}},\ \bibinfo {pages} {8577} (\bibinfo {year} {2006})}\BibitemShut {NoStop}%
	\bibitem [{\citenamefont {Mucha}\ \emph {et~al.}(2010)\citenamefont {Mucha}, \citenamefont {Richardson}, \citenamefont {Macon}, \citenamefont {Porter},\ and\ \citenamefont {Onnela}}]{mucha2010}%
	  \BibitemOpen
	  \bibfield  {author} {\bibinfo {author} {\bibfnamefont {P.~J.}\ \bibnamefont {Mucha}}, \bibinfo {author} {\bibfnamefont {T.}~\bibnamefont {Richardson}}, \bibinfo {author} {\bibfnamefont {K.}~\bibnamefont {Macon}}, \bibinfo {author} {\bibfnamefont {M.~A.}\ \bibnamefont {Porter}},\ and\ \bibinfo {author} {\bibfnamefont {J.-P.}\ \bibnamefont {Onnela}},\ }\bibfield  {title} {\bibinfo {title} {Community structure in time-dependent, multiscale, and multiplex networks},\ }\href@noop {} {\bibfield  {journal} {\bibinfo  {journal} {science}\ }\textbf {\bibinfo {volume} {328}},\ \bibinfo {pages} {876} (\bibinfo {year} {2010})}\BibitemShut {NoStop}%
	\bibitem [{\citenamefont {Betzel}\ and\ \citenamefont {Bassett}(2017)}]{BETZEL201773}%
	  \BibitemOpen
	  \bibfield  {author} {\bibinfo {author} {\bibfnamefont {R.~F.}\ \bibnamefont {Betzel}}\ and\ \bibinfo {author} {\bibfnamefont {D.~S.}\ \bibnamefont {Bassett}},\ }\bibfield  {title} {\bibinfo {title} {Multi-scale brain networks},\ }\href {https://doi.org/https://doi.org/10.1016/j.neuroimage.2016.11.006} {\bibfield  {journal} {\bibinfo  {journal} {NeuroImage}\ }\textbf {\bibinfo {volume} {160}},\ \bibinfo {pages} {73} (\bibinfo {year} {2017})},\ \bibinfo {note} {functional Architecture of the Brain}\BibitemShut {NoStop}%
	\bibitem [{\citenamefont {Garabed}\ \emph {et~al.}(2020)\citenamefont {Garabed}, \citenamefont {Jolles}, \citenamefont {Garira}, \citenamefont {Lanzas}, \citenamefont {Gutierrez},\ and\ \citenamefont {Rempala}}]{grja2020}%
	  \BibitemOpen
	  \bibfield  {author} {\bibinfo {author} {\bibfnamefont {R.~B.}\ \bibnamefont {Garabed}}, \bibinfo {author} {\bibfnamefont {A.}~\bibnamefont {Jolles}}, \bibinfo {author} {\bibfnamefont {W.}~\bibnamefont {Garira}}, \bibinfo {author} {\bibfnamefont {C.}~\bibnamefont {Lanzas}}, \bibinfo {author} {\bibfnamefont {J.}~\bibnamefont {Gutierrez}},\ and\ \bibinfo {author} {\bibfnamefont {G.}~\bibnamefont {Rempala}},\ }\bibfield  {title} {\bibinfo {title} {Multi-scale dynamics of infectious diseases},\ }\href {https://doi.org/10.1098/rsfs.2019.0118} {\bibfield  {journal} {\bibinfo  {journal} {Interface Focus}\ }\textbf {\bibinfo {volume} {10}},\ \bibinfo {pages} {20190118} (\bibinfo {year} {2020})}\BibitemShut {NoStop}%
	\bibitem [{\citenamefont {Brockmann}\ and\ \citenamefont {Helbing}(2013)}]{dbdh2013}%
	  \BibitemOpen
	  \bibfield  {author} {\bibinfo {author} {\bibfnamefont {D.}~\bibnamefont {Brockmann}}\ and\ \bibinfo {author} {\bibfnamefont {D.}~\bibnamefont {Helbing}},\ }\bibfield  {title} {\bibinfo {title} {The hidden geometry of complex, network-driven contagion phenomena},\ }\href {https://doi.org/10.1126/science.1245200} {\bibfield  {journal} {\bibinfo  {journal} {Science}\ }\textbf {\bibinfo {volume} {342}},\ \bibinfo {pages} {1337} (\bibinfo {year} {2013})}\BibitemShut {NoStop}%
	\bibitem [{\citenamefont {Quaranta}\ \emph {et~al.}(2020)\citenamefont {Quaranta}, \citenamefont {Formica}, \citenamefont {Machado}, \citenamefont {Lacarbonara},\ and\ \citenamefont {Masri}}]{qgfg2020}%
	  \BibitemOpen
	  \bibfield  {author} {\bibinfo {author} {\bibfnamefont {G.}~\bibnamefont {Quaranta}}, \bibinfo {author} {\bibfnamefont {G.}~\bibnamefont {Formica}}, \bibinfo {author} {\bibfnamefont {J.~T.}\ \bibnamefont {Machado}}, \bibinfo {author} {\bibfnamefont {W.}~\bibnamefont {Lacarbonara}},\ and\ \bibinfo {author} {\bibfnamefont {S.~F.}\ \bibnamefont {Masri}},\ }\bibfield  {title} {\bibinfo {title} {Understanding covid-19 nonlinear multi-scale dynamic spreading in italy},\ }\href@noop {} {\bibfield  {journal} {\bibinfo  {journal} {Nonlinear Dynamics}\ }\textbf {\bibinfo {volume} {101}},\ \bibinfo {pages} {1583} (\bibinfo {year} {2020})}\BibitemShut {NoStop}%
	\bibitem [{\citenamefont {Flack}(2012)}]{fljc2012}%
	  \BibitemOpen
	  \bibfield  {author} {\bibinfo {author} {\bibfnamefont {J.~C.}\ \bibnamefont {Flack}},\ }\bibfield  {title} {\bibinfo {title} {Multiple time-scales and the developmental dynamics of social systems},\ }\href {https://doi.org/10.1098/rstb.2011.0214} {\bibfield  {journal} {\bibinfo  {journal} {Philosophical Transactions of the Royal Society B: Biological Sciences}\ }\textbf {\bibinfo {volume} {367}},\ \bibinfo {pages} {1802} (\bibinfo {year} {2012})}\BibitemShut {NoStop}%
	\bibitem [{\citenamefont {Saram{\"a}ki}\ and\ \citenamefont {Moro}(2015)}]{saramaki2015}%
	  \BibitemOpen
	  \bibfield  {author} {\bibinfo {author} {\bibfnamefont {J.}~\bibnamefont {Saram{\"a}ki}}\ and\ \bibinfo {author} {\bibfnamefont {E.}~\bibnamefont {Moro}},\ }\bibfield  {title} {\bibinfo {title} {From seconds to months: an overview of multi-scale dynamics of mobile telephone calls},\ }\href@noop {} {\bibfield  {journal} {\bibinfo  {journal} {The European Physical Journal B}\ }\textbf {\bibinfo {volume} {88}},\ \bibinfo {pages} {1} (\bibinfo {year} {2015})}\BibitemShut {NoStop}%
	\bibitem [{\citenamefont {Guo}\ \emph {et~al.}(2016)\citenamefont {Guo}, \citenamefont {Wang}, \citenamefont {Li}, \citenamefont {Wang},\ and\ \citenamefont {Tang}}]{guo2016}%
	  \BibitemOpen
	  \bibfield  {author} {\bibinfo {author} {\bibfnamefont {C.}~\bibnamefont {Guo}}, \bibinfo {author} {\bibfnamefont {J.}~\bibnamefont {Wang}}, \bibinfo {author} {\bibfnamefont {J.}~\bibnamefont {Li}}, \bibinfo {author} {\bibfnamefont {Z.}~\bibnamefont {Wang}},\ and\ \bibinfo {author} {\bibfnamefont {S.}~\bibnamefont {Tang}},\ }\bibfield  {title} {\bibinfo {title} {Kinetic pathways and mechanisms of two-step nucleation in crystallization},\ }\href@noop {} {\bibfield  {journal} {\bibinfo  {journal} {The journal of physical chemistry letters}\ }\textbf {\bibinfo {volume} {7}},\ \bibinfo {pages} {5008} (\bibinfo {year} {2016})}\BibitemShut {NoStop}%
	\bibitem [{\citenamefont {Xu}\ \emph {et~al.}(2021)\citenamefont {Xu}, \citenamefont {Cao}, \citenamefont {Liu},\ and\ \citenamefont {Wang}}]{xu2021role}%
	  \BibitemOpen
	  \bibfield  {author} {\bibinfo {author} {\bibfnamefont {S.}~\bibnamefont {Xu}}, \bibinfo {author} {\bibfnamefont {D.}~\bibnamefont {Cao}}, \bibinfo {author} {\bibfnamefont {Y.}~\bibnamefont {Liu}},\ and\ \bibinfo {author} {\bibfnamefont {Y.}~\bibnamefont {Wang}},\ }\bibfield  {title} {\bibinfo {title} {Role of additives in crystal nucleation from solutions: A review},\ }\href@noop {} {\bibfield  {journal} {\bibinfo  {journal} {Crystal Growth \& Design}\ }\textbf {\bibinfo {volume} {22}},\ \bibinfo {pages} {2001} (\bibinfo {year} {2021})}\BibitemShut {NoStop}%
	\bibitem [{\citenamefont {Evans}\ and\ \citenamefont {Fu}(2018)}]{evans2018}%
	  \BibitemOpen
	  \bibfield  {author} {\bibinfo {author} {\bibfnamefont {T.}~\bibnamefont {Evans}}\ and\ \bibinfo {author} {\bibfnamefont {F.}~\bibnamefont {Fu}},\ }\bibfield  {title} {\bibinfo {title} {Opinion formation on dynamic networks: identifying conditions for the emergence of partisan echo chambers},\ }\href@noop {} {\bibfield  {journal} {\bibinfo  {journal} {Royal Society open science}\ }\textbf {\bibinfo {volume} {5}},\ \bibinfo {pages} {181122} (\bibinfo {year} {2018})}\BibitemShut {NoStop}%
	\bibitem [{\citenamefont {T{\'o}th}\ \emph {et~al.}(2021)\citenamefont {T{\'o}th}, \citenamefont {Wachs}, \citenamefont {Di~Clemente}, \citenamefont {Jakobi}, \citenamefont {S{\'a}gv{\'a}ri}, \citenamefont {Kert{\'e}sz},\ and\ \citenamefont {Lengyel}}]{toth2021}%
	  \BibitemOpen
	  \bibfield  {author} {\bibinfo {author} {\bibfnamefont {G.}~\bibnamefont {T{\'o}th}}, \bibinfo {author} {\bibfnamefont {J.}~\bibnamefont {Wachs}}, \bibinfo {author} {\bibfnamefont {R.}~\bibnamefont {Di~Clemente}}, \bibinfo {author} {\bibfnamefont {{\'A}.}~\bibnamefont {Jakobi}}, \bibinfo {author} {\bibfnamefont {B.}~\bibnamefont {S{\'a}gv{\'a}ri}}, \bibinfo {author} {\bibfnamefont {J.}~\bibnamefont {Kert{\'e}sz}},\ and\ \bibinfo {author} {\bibfnamefont {B.}~\bibnamefont {Lengyel}},\ }\bibfield  {title} {\bibinfo {title} {Inequality is rising where social network segregation interacts with urban topology},\ }\href@noop {} {\bibfield  {journal} {\bibinfo  {journal} {Nature communications}\ }\textbf {\bibinfo {volume} {12}},\ \bibinfo {pages} {1143} (\bibinfo {year} {2021})}\BibitemShut {NoStop}%
	\bibitem [{\citenamefont {Gottwald}(2015)}]{gwga2015}%
	  \BibitemOpen
	  \bibfield  {author} {\bibinfo {author} {\bibfnamefont {G.~A.}\ \bibnamefont {Gottwald}},\ }\bibfield  {title} {\bibinfo {title} {Model reduction for networks of coupled oscillators},\ }\href {https://doi.org/10.1063/1.4921295} {\bibfield  {journal} {\bibinfo  {journal} {Chaos: An Interdisciplinary Journal of Nonlinear Science}\ }\textbf {\bibinfo {volume} {25}},\ \bibinfo {pages} {053111} (\bibinfo {year} {2015})}\BibitemShut {NoStop}%
	\bibitem [{\citenamefont {Aoki}\ and\ \citenamefont {Aoyagi}(2009)}]{atat2009}%
	  \BibitemOpen
	  \bibfield  {author} {\bibinfo {author} {\bibfnamefont {T.}~\bibnamefont {Aoki}}\ and\ \bibinfo {author} {\bibfnamefont {T.}~\bibnamefont {Aoyagi}},\ }\bibfield  {title} {\bibinfo {title} {Co-evolution of phases and connection strengths in a network of phase oscillators},\ }\href {https://doi.org/10.1103/PhysRevLett.102.034101} {\bibfield  {journal} {\bibinfo  {journal} {Phys. Rev. Lett.}\ }\textbf {\bibinfo {volume} {102}},\ \bibinfo {pages} {034101} (\bibinfo {year} {2009})}\BibitemShut {NoStop}%
	\bibitem [{\citenamefont {Aoki}\ and\ \citenamefont {Aoyagi}(2011)}]{atat2011}%
	  \BibitemOpen
	  \bibfield  {author} {\bibinfo {author} {\bibfnamefont {T.}~\bibnamefont {Aoki}}\ and\ \bibinfo {author} {\bibfnamefont {T.}~\bibnamefont {Aoyagi}},\ }\bibfield  {title} {\bibinfo {title} {Self-organized network of phase oscillators coupled by activity-dependent interactions},\ }\href {https://doi.org/10.1103/PhysRevE.84.066109} {\bibfield  {journal} {\bibinfo  {journal} {Phys. Rev. E}\ }\textbf {\bibinfo {volume} {84}},\ \bibinfo {pages} {066109} (\bibinfo {year} {2011})}\BibitemShut {NoStop}%
	\bibitem [{\citenamefont {Hancock}\ and\ \citenamefont {Gottwald}(2018)}]{hancock2018}%
	  \BibitemOpen
	  \bibfield  {author} {\bibinfo {author} {\bibfnamefont {E.~J.}\ \bibnamefont {Hancock}}\ and\ \bibinfo {author} {\bibfnamefont {G.~A.}\ \bibnamefont {Gottwald}},\ }\bibfield  {title} {\bibinfo {title} {Model reduction for kuramoto models with complex topologies},\ }\href@noop {} {\bibfield  {journal} {\bibinfo  {journal} {Physical Review E}\ }\textbf {\bibinfo {volume} {98}},\ \bibinfo {pages} {012307} (\bibinfo {year} {2018})}\BibitemShut {NoStop}%
	\bibitem [{\citenamefont {Smith}\ and\ \citenamefont {Gottwald}(2020)}]{smith2020}%
	  \BibitemOpen
	  \bibfield  {author} {\bibinfo {author} {\bibfnamefont {L.~D.}\ \bibnamefont {Smith}}\ and\ \bibinfo {author} {\bibfnamefont {G.~A.}\ \bibnamefont {Gottwald}},\ }\bibfield  {title} {\bibinfo {title} {Model reduction for the collective dynamics of globally coupled oscillators: From finite networks to the thermodynamic limit},\ }\href@noop {} {\bibfield  {journal} {\bibinfo  {journal} {Chaos: An Interdisciplinary Journal of Nonlinear Science}\ }\textbf {\bibinfo {volume} {30}} (\bibinfo {year} {2020})}\BibitemShut {NoStop}%
	\bibitem [{\citenamefont {Stam}(2014)}]{stam2014}%
	  \BibitemOpen
	  \bibfield  {author} {\bibinfo {author} {\bibfnamefont {C.~J.}\ \bibnamefont {Stam}},\ }\bibfield  {title} {\bibinfo {title} {Modern network science of neurological disorders},\ }\href@noop {} {\bibfield  {journal} {\bibinfo  {journal} {Nature Reviews Neuroscience}\ }\textbf {\bibinfo {volume} {15}},\ \bibinfo {pages} {683} (\bibinfo {year} {2014})}\BibitemShut {NoStop}%
	\end{thebibliography}

%


\end{document}